\documentclass[11pt]{article}

\usepackage{amsfonts,amsmath,amssymb,verbatim,setspace, array,url}
\usepackage{enumitem}

\begin{document}

\newtheorem{theorem}{Theorem}[section]
\newtheorem{ex}[theorem]{Example}

\title{Decoding Generalized Reed-Solomon Codes and Its Application to RLCE Encryption Scheme}
\author{Yongge Wang\\ Department of SIS, UNC Charlotte, USA. \\
{yongge.wang@uncc.edu}}

\maketitle

\begin{abstract}
This paper compares the efficiency of various algorithms 
for implementing quantum resistant public key encryption scheme RLCE
on 64-bit CPUs. By optimizing various algorithms for polynomial 
and matrix operations over finite fields, we obtained several
interesting (or even surprising) results.
For example, it is well known (e.g., Moenck 1976 \cite{moenck1976practical}) that 
Karatsuba's algorithm outperforms classical polynomial multiplication
algorithm from the degree 15 and above (practically, Karatsuba's algorithm only
outperforms classical polynomial multiplication
algorithm from the degree 35 and above ). Our experiments
show that 64-bit optimized Karatsuba's algorithm will only 
outperform 64-bit optimized classical polynomial multiplication
algorithm for polynomials of degree 115 and above over finite field 
$GF(2^{10})$. The second interesting (surprising) result shows 
that 64-bit optimized Chien's search algorithm ourperforms all other 64-bit optimized 
polynomial root finding algorithms such as BTA and FFT for polynomials 
of all degrees over finite field $GF(2^{10})$.  The third interesting (surprising) 
result shows that 64-bit optimized  Strassen matrix multiplication algorithm
only outperforms 64-bit optimized  classical matrix multiplication algorithm
for matrices of dimension 750 and above over finite field 
$GF(2^{10})$. It should be noted that existing literatures
and practices recommend Strassen matrix multiplication 
algorithm for matrices of dimension 40 and above.
All our experiments are done on a 64-bit MacBook Pro with i7 CPU and single thread C codes.
It should be noted that the reported results should be 
appliable to 64 or larger bits CPU architectures. For 32 or smaller bits CPUs, these results may not 
be applicable. The source code
and library for the algorithms covered in this paper are available at
\url{http://quantumca.org/}. 
\end{abstract}


\section{Introduction}
This paper investigates efficient algorithms for implementing 
quantum resistant public key encryption scheme RLCE.
Specifically, we will compare various decoding algorithms
for generalized Reed-Solomon (GRS) codes: Berlekamp-Massey decoding algorithms; 
Berlekamp-Welch decoding algorithms; Euclidean decoding algorithms; and 
list decoding algorithm. The paper also compares
various efficient algorithms for polynomial and matrix operations over finite 
fields. For example, the paper will cover Chien's search algorithm; Berlekamp trace algorithm;  
Forney's algorithm, Strassen algorithm, and many others. 
The focus of this document is to identify the optimized algorithms
for implementing the RLCE encryption scheme by Wang \cite{7541753,wangrlcelong}
on 64-bit CPUs. The experimental results for these algorithms over finite fields $GF(2^{10})$ and  $GF(2^{11})$
are reported in this document.  

\section{Finite fields}
\subsection{Representation of elements in finite fields}
In this section, we present a Layman's guide to several representations of elements 
in a finite field $GF(q)$. We assume that the reader is familiar with the finite 
field $GF(p)=Z_p$ for a prime number $p$ and we concentrate on 
the construction of finite fields $GF(p^m)$.

\noindent
{\bf Polynomials:} Let $\pi(x)$ be an irreducible polynomial of degree $m$ 
over $GF(p)$. Then the set of all polynomials in $x$ of degree $\le m-1$ and coefficients
from $GF(p)$ form the finite field $GF(p^m)$ where field elements addition and multiplication
are defined as polynomial addition and multiplication modulo $\pi(x)$

For an irreducible polynomial $f(x)\in GF(p)[x]$ of degree $m$, $f(x)$ has 
a root $\alpha$ in $GF(p^m)$. Furthermore,
all roots of $f(x)$ are given by the $m$ distinct elements $\alpha, \alpha^p, \cdots, \alpha^{p^{m-1}}\in GF(p^m)$.

\noindent
{\bf Generator and primitive polynomial:}
A primitive polynomial $\pi(x)$ of degree $m$ over $GF(p)$ is an
irreducible polynomial  that has a root $\alpha$ in $GF(p^m)$ 
so that $GF(p^m)=\{0\}\cup \{\alpha^i: i=0,\cdots, p^m-1\}$. As an example for $GF(2^3)$,  
$x^3+x+1$ is a primitive polynomial with root $\alpha=010$. That is,
$$\begin{array}{|l|l|l|l|}\hline
\alpha^0=001 & \alpha^1=010 & \alpha^2=100 &\alpha^3=011\\ \hline
\alpha^4=110 & \alpha^5=111 & \alpha^6=101 &\alpha^7=001\\ \hline
\end{array}
$$
Note that not all irreducible polynomials are primitive. For example $1+x+x^2+x^3+x^4$
is irreducible over $GF(2)$ but not primitive. 
The root of a generator polynomial is called a primitive element.

\noindent
{\bf Matrix approach:} The companion matrix of a polynomial $\pi(x)=a_0+a_1x+\cdots+a_{m-1}x^{m-1}+x^m$
is defined to be the $m\times m$ matrix
$$M=\left(
\begin{array}{ccccc}
0&1&0 &\cdots &0\\
0&0&1 &\cdots &0\\
\vdots&\vdots&\vdots&\ddots &\vdots\\
0&0&0 &\cdots &1\\
-a_0&-a_1&-a_2 &\cdots &-a_{m-1}\\
\end{array}
\right)
$$
The set of matrices $0,M, \cdots, M^{p^m-1}$  with matrix addition and multiplication 
over $GF(p)$ forms the finite field $GF(p^m)$. 

\noindent
{\bf Splitting field:} Let $\pi(x)\in GF(p)[x]$ be a degree $m$ irreducible polynomial. Then $GF(p^m)$
can be considered as a splitting field of $\pi(x)$ over $GF(p)$.  That is, assume that 
$\pi(x)=(x-\alpha_1)\cdots (x-\alpha_m)$ in  $GF(p^m)$. Then $GF(p^m)$
is obtained by adjoining these algebraic elements $\alpha_1, \cdots, \alpha_m$ to  $GF(p)$. 

\subsection{Finite field arithmetic}
Let $\alpha$ be a primitive element in $GF(q)$. Then for 
each non-zero $x\in GF(q)$, there exists a $0\le y\le q-2$ such that $x=\alpha^y$ where
$y$ is called the discrete logarithm of $x$. When field elements are represented using their 
discrete logarithms, multiplication and division are efficient since they are reduced 
to integer addition and subtraction modulo $q-1$. For additions,  one may use Zech's logarithm
which is defined as 
\begin{equation}
Z(y): y\mapsto \log_\alpha(1+\alpha^y).
\end{equation}
That is, for a field element $\alpha^y$, we have 
$\alpha^{Z(y)}=1+\alpha^y$.
If one stores Zech's logarithm in a table as pairs $(y, Z(y))$,  then the addition could be calculated as
$$\alpha^{y_1}+\alpha^{y_2}=\alpha^{y_1}(1+\alpha^{y_2-y_1})=
\alpha^{y_1}\alpha^{Z(y_2-y_1)}=\alpha^{y_1+Z(y_2-y_1)}.$$
For the finite field $GF(2^m)$, the addition is the efficient XOR operation. Thus it is 
better to store two tables to speed up the multiplication: discrete logarithm table 
and exponentiation tables. For the discrete logarithm table, one obtains 
$y$ on input $x$ such that $x=\alpha^y$. For the exponentiation table,
one obtains $y$ on input $x$ such that $y=\alpha^x$.
In order to multiply two field elements $x_1, x_2$, one first gets their discrete logarithms
$y_1,y_2$ respectively. Then one calculates $y=y_1+y_2$.
Next one looks up the exponentiation table to find out the value of $\alpha^{y}$.
Note that we have $x_1x_2=\alpha^{y_1}\alpha^{y_2}=\alpha^{y_1+y_2}$.

\section{Polynomial and matrix arithmetic}
\subsection{Fast Fourier Transform (FFT)}
The Fast Fourier transform maps a polynomial $f(x)=f_0+f_1x+\cdots + f_{n-1}x^{n-1}$
to its values 
$$\mbox{FFT}(f(x))=(f(\alpha^0), \cdots, f(\alpha^{n-1})).$$ 
Fast Fourier Transforms (FFT) are useful for improving RLCE decryption performance.
In this section, we review FFT over $GF(p^m)$ with $p>2$ and FFT over $GF(2^m)$.
The applications of FFTs will be presented in next sections.

\subsubsection{FFT over $GF(p^m)$ with $p>2$}
Let $n$ be even and $\alpha$ be a primitive $n$th root of unit in $GF(p^m)$ with $p>2$.
That is, $\alpha^n=1$. It should be noted that for a field with characteristics 2 such as $GF(2^m)$, such kind 
of primitive roots do not exist.  FFT uses the fact that 
$$(\alpha^{i})^2=(\alpha^{i+\frac{n}{2}})^2$$
for all $i$. Note that for the complex number based FFT, this fact is equivalent to the 
fact that $\alpha^{\frac{n}{2}}=-1$ though the value ``$-1$'' should be interpreted appropriately in finite fields.
Suppose that $f(x)=f_0+f_1x+\cdots + f_{n-1}x^{n-1}$.
If $n$ is odd, we can add an term $0\cdot x^{n-1}$ to $f(x)$ so that $f(x)$ has degree $n-1$.
Define  the even index polynomial
$f^{[0]}(x)=\sum_{i=0}^{\frac{n-2}{2}}f_{2i}x^{i}$ 
and the odd index polynomial $f^{[1]}(x)=\sum_{i=0}^{\frac{n-2}{2}}f_{2i+1}x^{i}$ of 
degree $\frac{n-2}{2}$. Since $f(x)=f^{[0]}(x^2)+xf^{[1]}(x^2)$, we can evaluate 
$f(x)$ on the $n$ points $\alpha^0, \cdots, \alpha^{n-1}$ by evaluating the two polynomials $f^{[0]}(x)$ and 
$f^{[1]}(x)$ on the $\frac{n}{2}$ points $\left\{\alpha^0, \alpha^2, \alpha^4, \cdots, \alpha^{2n-2}\right\}=
\left\{\alpha^0, \alpha^2, \alpha^4, \cdots, \alpha^{\frac{n}{2}-1}\right\}$ and then 
combining  the results. 
By carrying out this process recursively, we can compute 
$\mbox{FFT}(f(x))$ in $O(n\log n)$ steps instead of $O(n^2)$ steps.

\subsubsection{FFT over $GF(2^m)$ and Cantor's algorithm}
\label{fft2m}
For finite fields with characteristics 2 such as $GF(2^m)$, one may use 
Cantor's algorithm \cite{cantor1989arithmetical} and its variants 
\cite{von1996arithmetic,gao2010additive} for efficient FFT computation.
These techniques are also called additive FFT algorithms 
and  could be used to compute $\mbox{FFT}(f(x))$ over $GF(2^m)$ in $O(m^22^m)$ steps. 

Let $\beta_0, \cdots, \beta_{d-1}\in GF(2^m)$ be linearly independent over $GF(2)$ and 
let $B$ be a subspace spanned by 
$\beta_i$'s over $GF(2)$. That is,
$$B=\mbox{span}(\beta_0,\cdots, \beta_{d-1})=\left\{\sum_{i=0}^{d-1}a_i\beta_i: a_i\in GF(2)\right\}.$$
For $0\le i<2^d$ with the binary representation $i=a_{d-1}a_{d-1}\cdots a_0$,  
the $i$-th element in $B$ is $B[i]=\sum_{i=0}^{d-1}a_i\beta_i$. 
For $0\le i\le d-1$, let $W_i=\mbox{span}(\beta_0,\cdots, \beta_{i})$. Then we have 
$$\{0\}=W_{-1}\subsetneq W_0\subsetneq W_1\subsetneq \cdots \subsetneq W_{d-1}$$
and $W_{i}=\left(\beta_{i}+W_{i-1}\right)\cup W_i$ for $i=0, \cdots, d-1$.
This can be further generalized to 
$$\beta+W_{i}=\left(\beta+\beta_{i}+W_{i-1}\right)\cup \left(\beta+W_i\right)$$
for $i=0, \cdots, d-1$ and all $\beta\in GF(2^m)$. Next define the minimal
polynomial $s_i(x)\in GF(2^m)[x]$ of $W_i$ as
$$s_i(x)=\prod_{\alpha\in W_i}(x-\alpha)$$
for $i=0, \cdots, d-1$. It is shown in \cite{von1996arithmetic} that 
$s_i(x)$ is a $GF(2)$-linearized polynomial where the concept of linearized polynomial is given in
Section \ref{linearsec}. Furthermore, by the fact that 
$$s_{i}(x)=\prod_{\alpha\in W_i}(x-\alpha)=\left(\prod_{\alpha\in W_{i-1}}(x-\alpha)\right)\left(
\prod_{\alpha\in \beta_i+W_{i-1}}(x-\alpha)\right)=s_{i-1}(x)\cdot s_{i-1}(x-\beta_i)$$
and by the fact that $s_i(x)$ is a linearized polynomial, we have 
$$s_i(x)=s_{i-1}(x)\cdot s_{i-1}(x-\beta_i)=s_{i-1}(x)\left(s_{i-1}(x)-s_{i-1}(\beta_i)\right)$$
for $i=0, \cdots, d-1$.
Table \ref{sig10} lists the polynomials $s_i(x)$ over $GF(2^{10})$ for the base 
$\beta_i=b_{9}b_8\cdots b_0$ where $b_j=0$ for $j\not=i$ and $b_i=1$.

\begin{table}[h]
\caption{Linearized polynomials $s_i(x)$ over $GF(2^{10})$}
\label{sig10}
\begin{center}
$\begin{array}{ll}
s_0(x)& =x^2+x\\ 
s_1(x)&=x^4+{\tt 0x007} x^2 +{\tt 0x006} x\\
s_2(x)&=x^8+{\tt 0x17d}  x^4++{\tt 0x205} x^2 +{\tt 0x379} x\\
s_3(x)&=x^{16}+{\tt 0x2b5} x^8+{\tt 0x3f4}  x^4+{\tt 0x177} x^2 +{\tt 0x037} x\\
s_4(x)&=x^{32}+{\tt 0x18a}x^{16}+{\tt 0x139} x^8+{\tt 0x353}  x^4+{\tt 0x3f4} x^2 +{\tt 0x015} x\\
s_5(x)&=x^{64}+{\tt 0x179}x^{32}+{\tt 0x0b3}x^{16}+{\tt 0x303} x^8+{\tt 0x09f}  x^4+{\tt 0x0b2} x^2 +{\tt 0x2e5} x\\
s_6(x)&=x^{128}+{\tt 0x394}x^{64}+{\tt 0x35f}x^{32}+{\tt 0x28f}x^{16}+{\tt 0x3ef} x^8+{\tt 0x041}  x^4+{\tt 0x0de} x^2 \\
&\quad +{\tt 0x135} x\\
s_7(x)&=x^{256}+{\tt 0x2bd}x^{128}+{\tt 0x2cf}x^{64}+{\tt 0x2e1}x^{32}+{\tt 0x1a5}x^{16}+{\tt 0x3f4} x^8+{\tt 0x279}  x^4\\
&\quad +{\tt 0x3a8} x^2 +{\tt 0x112} x\\
s_8(x)&=x^{512}+{\tt 0x214}x^{256}+{\tt 0x043}x^{128}+{\tt 0x292}x^{64}+{\tt 0x070}x^{32}+{\tt 0x0ce}x^{16}+{\tt 0x0b3} x^8\\
&\quad +{\tt 0x24c}  x^4+{\tt 0x081} x^2 +{\tt 0x204} x
\end{array}$
\end{center}
\end{table}

Table \ref{sig11} lists the polynomials $s_i(x)$ over $GF(2^{10})$ for the base 
$\beta_i=b_{10}b_9\cdots b_0$ where $b_j=0$ for $j\not=i$ and $b_i=1$.
\begin{table}[h]
\caption{Linearized polynomials $s_i(x)$ over $GF(2^{11})$}
\label{sig11}
\begin{center}
$\begin{array}{ll}
s_0(x)& =x^2+x\\ 
s_1(x)&=x^4+{\tt 0x007} x^2 +{\tt 0x006} x\\
s_2(x)&=x^8+{\tt 0x17d}  x^4++{\tt 0x60c} x^2 +{\tt 0x770} x\\
s_3(x)&=x^{16}+{\tt 0x4c3} x^8+{\tt 0x6c0}  x^4++{\tt 0x390} x^2 +{\tt 0x192} x\\
s_4(x)&=x^{32}+{\tt 0x48a}x^{16}+{\tt 0x278} x^8+{\tt 0x528}  x^4+{\tt 0x274} x^2 +{\tt 0x1af} x\\
s_5(x)&=x^{64}+{\tt 0x69e}x^{32}+{\tt 0x4ec}x^{16}+{\tt 0x619} x^8+{\tt 0x4fd}  x^4+{\tt 0x05b} x^2 \\
&\quad+{\tt 0x0cc} x\\
s_6(x)&=x^{128}+{\tt 0x734}x^{64}+{\tt 0x294}x^{32}+{\tt 0x357}x^{16}+{\tt 0x4a0} x^8+{\tt 0x1f8}  x^4\\
&\quad+{\tt 0x211} x^2 +{\tt 0x1bf} x\\
s_7(x)&=x^{256}+{\tt 0x50b}x^{128}+{\tt 0x52b}x^{64}+{\tt 0x31b}x^{32}+{\tt 0x0da}x^{16}+{\tt 0x56e} x^8\\
&\quad+{\tt 0x0c0}  x^4+{\tt 0x230} x^2 +{\tt 0x47e} x\\
s_8(x)&=x^{512}+{\tt 0x385}x^{256}+{\tt 0x584}x^{128}+{\tt 0x4b0}x^{64}+{\tt 0x11f}x^{32}+{\tt 0x2ef}x^{16}\\
&\quad+{\tt 0x261} x^8+{\tt 0x429}  x^4+{\tt 0x68d} x^2 +{\tt 0x185} x\\
s_9(x)&=x^{1024}+{\tt 0x703}x^{512}+{\tt 0x781}x^{256}+{\tt 0x7c9}x^{128}+{\tt 0x7da}x^{64}+{\tt 0x4d2}x^{32}\\
&\quad+{\tt 0x444}x^{16}+{\tt 0x60c} x^8+{\tt 0x69f}  x^4+{\tt 0x5d7} x^2 +{\tt 0x542} x\\
\end{array}$
\end{center}
\end{table}

With these preliminary definition, we first review von zur Gathen and Gerhard's additive FFT algorithm.
Let $\beta_0, \cdots, \beta_{d-1}\in GF(2^m)$ be linearly independent over $GF(2)$ and 
let $B=\mbox{span}(\beta_0,\cdots, \beta_{d-1})$. 
For a given polynomial $f(x)$ of degree less than $2^{d}$,
we evaluate $f(x)$ over all points in $B$ using the following 
algorithm ${\tt GGFFT}(f(x), d, B)=\langle f(B[0]), \cdots, f(B[2^d-1])\rangle$.
The algorithm assumes that the polynomials $s_i(x)$, the values $s_i(\beta)$ and $s_i(\beta_{i+1})^{-1}$
for $-1\le i<j\le d-1$ are pre-computed.

\vskip 5pt
\noindent
{\bf Gathen-Gerhard's} ${\tt GGFFT}(f(x), i,d, B, b_{i+1}, \cdots, b_{d-1})$:

\noindent
{\tt Input:} $i\in [-1, d-1]$, $f\in GF(2^m)[x]$, $\deg(f(x))<2^{i+1}$, and $b_{i+1}, \cdots, b_{d-1}\in GF(2)$.

\noindent
{\tt Output:} $\langle f(\alpha+\beta):\alpha\in W_i\rangle$ where $\beta=b_{i+1}\beta_{i+1}+\cdots+b_{d-1}\beta_{d-1}$.

\noindent
{\tt Algorithm:} 
\begin{enumerate}
\item If $i=-1$, return $f$.
\item Compute $g(x), r_0(x)\in GF(2^m)[x]$ such that 
$$f(x)=g(x)\left(s_{i-1}(x)+s_{i-1}(\beta)\right) +r_0(x)\mbox{ and }\deg(r_0(x))<2^{i-1}.$$
Let $r_1(x)=r_0(x)+s_{i-1}(\beta_i)\cdot g(x)$.
\item Return ${\tt GGFFT}(r_0(x), i-1,d, B, 0, b_{i+1}, \cdots, b_{d-1}) \cup {\tt GGFFT}(r_1(x), i-1,d, B, 1, b_{i+1}, \cdots, b_{d-1})$.
\end{enumerate}

It is shown in \cite{von1996arithmetic} that the algorithm ${\tt GGFFT}(f(x), d, B)$ runs 
with $O(2^dd^2)$ multiplications and additions.
We next review Gao-Mateer's FFT algorithm \cite{gao2010additive} which runs with 
$O(2^dd)$ multiplications and $O(2^dd^2)$ additions.

\vskip 5pt
\noindent
{\bf Gao-Mateer's} ${\tt GMFFT}(f(x), d, B))$: 

\noindent
{\tt Input:} $f\in GF(2^m)[x]$, $\deg(f(x))<2^{d}$, $B=\mbox{span}(\beta_0,\cdots, \beta_{d-1})$

\noindent
{\tt Output:} $\langle f(B[0]), \cdots, f(B[2^d-1])\rangle$. 

\noindent
{\tt Algorithm:} 
\begin{enumerate}
\item If $\deg(f(x))=0$, return $\langle f(0), f(0)\rangle$.
\item If $d=1$, return $\langle f(0), f(\beta_1)\rangle$.
\item Let $g(x)=f(\beta_dx)$.
\item Use the algorithm in the next paragraph to compute ${\tt Taylor}(g(x))$ as in (\ref{taylorg}) and let 
\begin{equation}
\label{g0g1}
g_0(x)=\sum_{i=0}^{l-1}g_{i,0}x^i \quad\mbox{ and }\quad g_1(x)=\sum_{i=0}^{l-1}g_{i,1}x^i.
\end{equation}
\item Let $\gamma_i=\beta_i\beta_d^{-1}$ and $\delta_i=\gamma_i^2-\gamma_i$ for $0\le i\le d-2$.
\item Let $G=\mbox{span}(\gamma_0,\cdots, \gamma_{d-2})$ and $D=\mbox{span}(\delta_0,\cdots, \delta_{d-2})$
\item Let 
$$\begin{array}{l}
{\tt FFT}(g_0(x), d-1, D)=\langle u_0, \cdots, u_{2^{d-1}-1}\rangle\\
{\tt FFT}(g_1(x), d-1, D)=\langle v_0, \cdots, v_{2^{d-1}-1}\rangle\\
\end{array}$$
\item Let $w_i=u_i+G[i]\cdot v_i$ and $w_{2^{d-1}+i}=w_i+v_i$ for $0\le i<2^{d-1}$.
\item Return $\langle  w_0, \cdots, w_{2^d-1}\rangle$.
\end{enumerate}

For a polynomial $g(x)$ of degree $2l-1$ over $GF(2^m)$, the Taylor expansion of $g(x)$ at $x^2-x$ is 
a list $\langle g_{0,0}+g_{0,1}x, \cdots, g_{l-1, 0}+g_{l-1,1}x\rangle$ where
\begin{equation}
\label{taylorg}
g(x)=(g_{0,0}+g_{0,1}x)+(g_{1,0}+g_{1,1}x)(x^2-x)+\cdots + (g_{l-1, 0}+g_{l-1,1}x)(x^2-x)^{l-1}
\end{equation}
and $g_{i,j}\in GF(2^m)$.  The Taylor expansion of $g(x)$ could be computed using the following algorithm ${\tt Taylor}(g(x))$:
\begin{enumerate}
\item If $\deg(g(x))< 2$, return $g(x)$.
\item Find $l$ such that $2^{l+1}<1+\deg(g(x)) \le 2^{l+2}$.
\item Let $g(x)=h_0(x)+x^{2^{l+1}}\left(h_1(x)+x^{2^l}h_2(x)\right)$
where $\deg(h_0)<2^{l+1}, \deg(h_1)<2^l, \deg(h_2)<2^l$.
\item Return $\langle  {\tt Taylor}(h_0(x)+x^{2^l}(h_1(x)+h_2(x))), {\tt Taylor}(h_1(x)+h_2(x)+x^{2^l}h_2(x))\rangle$.
\end{enumerate}

It is shown in \cite{gao2010additive}  that the  algorithm ${\tt GMFFT}$ uses
at most $2^{d-1}\log^2(2^d)$ additions and $2^{d+1}\log(2^d)$ multiplications.

\subsubsection{Inverse FFT over $GF(p^m)$}
\label{ifftpm}
For a polynomial $f(x)=f_0+f_1x+\cdots + f_{n-1}x^{n-1}$,
the Inverse FFT is defined as 
$$\mbox{IFFT}(\mbox{FFT}(f(x)))=\mbox{IFFT}(f(\alpha^0), \cdots, f(\alpha^{n-1}))=(f_0,\cdots, f_{n-1}).$$
Assume that $n=p^m-1$ and $\alpha^n=1$.
The Mattson-Solomon polynomial of $f$ is defined as
\begin{equation}
\label{invFFT}
F(x)=\sum_{i=0}^{n-1}f(\alpha^i)x^{n-i}.
\end{equation}
By the fact that 
$$x^n-1=(x-1)(1+x+\cdots+x^{n-1}),$$
we have 
$\displaystyle\sum_{i=0}^{n-1}a^i=0$ for all $a\in GF(q)$ with $a\not=1$. Then 
\begin{equation}
\label{fdfdfghifft}
\begin{array}{ll}
F(\alpha^j)&=\displaystyle\sum_{i=0}^{n-1}f(\alpha^i)\alpha^{j(n-i)}\\
&=\displaystyle\sum_{i=0}^{n-1}\displaystyle\sum_{u=0}^{n-1}f_u\alpha^{ui}\alpha^{j(n-i)}\\
&=\displaystyle\sum_{u=0}^{n-1}  f_u \displaystyle\sum_{i=0}^{n-1}\alpha^{(u-j)i}\\
&=nf_j \\
\end{array}
\end{equation}
It follows that $\mbox{IFFT}(\mbox{FFT}(f(x)))=\mbox{FFT}\left(\frac{F(x)}{n}\right)$.

The relationship between FFT and IFFT may also be explained using the fact 
for Vendermonde matrix that $V_n(\alpha^{0}, \cdots, \alpha^{n-1})^{-1}=
\frac{V_n(\alpha^{-0}, \cdots, \alpha^{-(n-1)})}{n}$. It is noted that 
$$\mbox{FFT}(f(x))=(f_0, \cdots, f_{n-1})\left(
\begin{array}{cccc}
1 & 1 &\cdots & 1\\
1 & \alpha^1 & \cdots &\alpha^{n-1}\\
1 & \alpha^2 & \cdots &\alpha^{2(n-1)}\\
\vdots&\vdots& \ddots &\vdots\\
1 & \alpha^{n-1}&\cdots &\alpha^{(n-1)^2}
\end{array}\right)=(f_0, \cdots, f_{n-1}) V_n(\alpha^0, \cdots, \alpha^{n-1})$$
On the other hand,
$$\begin{array}{ll}
\mbox{FFT}(F(x))&=\left(f(\alpha^0), \cdots, f(\alpha^{n-1})\right)\left(
\begin{array}{cccc}
1 & \alpha^{n} &\cdots & \alpha^{n(n-1)}\\
1 & \alpha^{n-1} & \cdots &\alpha^{(n-1)(n-1)}\\
1 & \alpha^{n-2}& \cdots &\alpha^{(n-1)(n-2)}\\
\vdots&\vdots& \ddots &\vdots\\
1 & \alpha^{1} & \cdots &\alpha^{(n-1)}\\
\end{array}\right)\\
&=\mbox{FFT}(f(x))\left(
\begin{array}{cccc}
1 & 1 &\cdots & 1\\
1 & \alpha^{-1}& \cdots &\alpha^{-(n-1)}\\
1 & \alpha^{-2} & \cdots &\alpha^{-(n-2)}\\
\vdots&\vdots& \ddots &\vdots\\
1 & \alpha^{-(n-1)} & \cdots &\alpha^{-(n-1)^2}\\
\end{array}\right)\\
&=\mbox{FFT}(f(x))\cdot V_n(\alpha^{-0}, \cdots, \alpha^{-(n-1)})\\
&=n\cdot \mbox{FFT}(f(x))\cdot V_n(\alpha^{0}, \cdots, \alpha^{n-1})^{-1}\\
&=n\cdot (f_0, \cdots, f_{n-1})
\end{array}$$

\subsubsection{Inverse FFT over $GF(2^m)$}
For FFT over $GF(2^m)$ in Section \ref{fft2m}, the output is in the order
$f(B[0]), \cdots, f(B[2^m-1])$ instead of the order $f(\alpha^0), \cdots, f(\alpha^{2^m-1})$.
Thus in order to calculate $F(x)$ in Section \ref{ifftpm}, we need to find a list of indices $j_0, \cdots, j_{2^{m-1}-1}$ such
that $B[j_i]=\alpha^i$ for $0\le i\le 2^{m-1}-1$. Then we can let 
$$F(x)=\sum_{i=0}^{n-1}f(B[j_i])x^{n-i}.$$
Similarly, after $\mbox{IFFT}(F(x))=(F(B[0]), \cdots, F(B[2^m-1]))$ is obtained, 
we will have $f_i={F(B[j_i])}$ for $0\le i\le 2^{m-1}-1$.
On the other hand, in order to use the techniques in Sections \ref{ifftpm}
and \ref{fft2m} to interpolate a polynomial, one essentially needs a base 
$\{\beta_0, \cdots, \beta_{m-1}\}$ to generate the entire field $GF(2^m)$
and to compute FFT over the entire field $GF(2^m)$.
This is inefficient for polynomials whose degrees are much smaller than $2^{m-1}$.

In the following, we describe the Chinese Reaminder Theorem based IFFT algorithm from von zur Gathen 
and Gerhard \cite{von1996arithmetic} that takes advantage of the additive FFT property.
Let $\beta_0, \cdots, \beta_{d-1}\in GF(2^m)$ be linearly independent over $GF(2)$ and 
let $B=\mbox{span}(\beta_0,\cdots, \beta_{d-1})$. 

\vskip 5pt
\noindent
{\bf Gathen-Gerhard's} ${\tt GGIFFT}(i,B, \beta, f(\beta+W_i))$:

\noindent
{\tt Input:} $i\in [0, d-1]$, $\beta$,
and $\langle f(\beta+W_i[0]), \cdots, f(\beta+W_i[{2^{i+1}-1}])\rangle$ where $\beta=\displaystyle\sum_{j=i+1}^{d-1}b_j\beta_j$ for some $b_{i+1}, \cdots, b_{d-1}\in GF(2)$.

\noindent
{\tt Output:} $f(x)\in GF(2^m)[x]$ with $\deg(f(x))<2^{i+1}$.

\noindent
{\tt Algorithm:} 
\begin{enumerate}
\item If $i=0$, then return $f(x)=\beta_0^{-1}(f(\beta)+f(\beta+\beta_0))x+f(\beta)+\beta_0^{-1}\beta(f(\beta)+f(\beta+\beta_0))$.

\item Let $\beta'=\beta+\beta_i$ and 
$$\begin{array}{l}
f_0(x)={\tt GGIFFT}(i-1, B, \beta, f(\beta+W_{i-1}))\\
f_1(x)={\tt GGIFFT}(i-1, B, \beta', f(\beta'+W_{i-1}))
\end{array}$$
where $\deg(f_0(x))<2^i$ and $\deg(f_1(x))<2^i$.
\item Return $f(x)=\left(s_{i-1}(x)+s_{i-1}(\beta)\right)\cdot (f_0(x)+f_1(x))\cdot s_{i-1}(\beta_i)^{-1}+f_0(x)$.
\end{enumerate}

\subsection{Polynomial multiplication I: Karatsuba algorithm}
For two polynomials $f(x)$ and $g(x)$, we can rewrite them as 
$$f(x)=f_1(x) x^{n_1} +f_2(x)\quad\quad\mbox{and}\quad\quad g(x)=g_1(x) x^{n_1} +g_2(x)$$
where $f_1, f_2, g_1,g_2$ has degree  less  than $n_1$. Then
$$f(x)g(x)=h_1(x)x^{2n_1}+h_2(x)x^{n_1}+h_3(x)$$
where 
$$\begin{array}{l}
h_1(x)=f_1(x)g_1(x)\\
h_2(x)=(f_1(x)+f_2(x))(g_1(x)+g_2(x))-h_1(x)-h_3(x)\\
h_3(x)=f_2(x)g_2(x)
\end{array}$$
Karatsuba's algorithm could be recursively called and the time complexity is $O(n^{1.59})$.
Our experiments show that Karatsuba's algorithm could improve the efficiency of RLCE scheme 
for most security parameters.

\subsection{Polynomial multiplication II:  FFT}
For RLCE  over $GF(p^m)$, one can use FFT
to speed up the polynomial multiplication and division.  
For two polynomials $f(x)$ and $g(x)$, we first compute
$\mbox{FFT}(f(x))$ and $\mbox{FFT}(g(x))$ in at most $O(n\log^2 n)$ steps. With  $n$ more multiplications,
we obtain $\mbox{FFT}(f(x)g(x))$. From $\mbox{FFT}(f(x)g(x))$, the interpolation can be computed
using the inverse FFT as $f(x)g(x)=\mbox{FFT}^{-1}(f(x)g(x))$.  This can be done in $O(n\log^2 n)$ steps.
Thus polynomial multiplication can be done in $O(n\log^2 n)$ steps.  Our experiements show
that FFT based polynomial multiplication helps none of the RLCE encryption schemes.

\subsection{Polynomial division}
Given polynomials $f(x)$ and $g(x)$ with $\deg(f)=n$ and $\deg(g)=n_1$, we want to find $q(x)$ and $r(x)$ such that 
$f(x)=g(x)q(x)+r(x)$ in $O(n\log n)$ step. The algorithm is described in terms of 
polynomials with infinite degrees which is called polynomial series. A polynomial with an infinite degree
has an inverse if it is in the form of $a_0+xh(x)$ where $a_0\not=0$ and $h(x)$ is a polynomial series. Furthermore, 
we have 
$(1+x)^{-1}=\sum_0^{\infty}(-x)^i$ and $\left(\sum_i^\infty (i+1)x^i\right)^{-1}=(1-x)^2$.
If we substitute $x$ with $\frac{1}{y}$ in $f(x)=g(x)q(x)+r(x)$, we obtain 
\begin{equation}
\label{fdsfd}
f^R(y)=q^R(y)g^R(y)+y^{n-n_1-1}r^R(y)=g^R(y)q^R(y) \mod y^{n-n_1-1}
\end{equation}
where $h^R(y)=y^{\deg(h)}h(\frac{1}{y})$ with the reversed order of coefficients for any polynomial $h$.
By the assumption that $g(x)$ has degree $n_1$, we know that $g^R$ is inevitable in the polynomial series.
Thus (\ref{fdsfd})  implies that 
\begin{equation}
\label{fgafdv}
q^R(y)=f^R(y) (g^R(y))^{-1} \mod y^{n-n_1-1}
\end{equation}
In order to compute  $q^R(y)$, only $n-n_1-1$ terms from the polynomial series $(g^R(y))^{-1}$ is required.
The following algorithm INV$(h(x), t)$ can be used to compute the first $t$ terms 
of $(h(x))^{-1}$ for $h(x)=\sum_{i=0}^{n_1-1}a_ix^i$.
\begin{enumerate}
\item If $t=1$, output $\frac{1}{a_0}$.
\item $h'=\mbox{INV}(h(x), \left\lceil \frac{t}{2}\right\rceil)$.
\item output $(h'(x)-(h(x)h'(x)-1)h'(x)) \mod x^t$.
\end{enumerate}
If the fast polynomial multiplication algorithm is used for the computation of 
$h'(x)-(h(x)h'(x)-1)h'(x)$, the the above algorithm INV$(h(x), t)$  uses $O(n_1\log n_1)$ steps.
The following is the $O(n\log n)$ algorithm for computing $q(x)$ and $r(x)$ given $f(x)$ and $g(x)$.
\begin{enumerate}
\item Let $f^R(x)=x^nf(\frac{1}{x})$ and $g^R(x)=x^{n_1}g(\frac{1}{x})$.
\item Let $(g^R(x))^{-1}(y)=\mbox{INV}(g^R(x), n-n_1-1)$.
\item Let $q^R(x)=f^R(x)(g^R(x))^{-1}(y)\mod x^{n-n_1-1}$.
\item Let $q(x)=x^{n-n_1-1}q^R(\frac{1}{x})$.
\item Let $r(x)=f(x)-q(x)g(x)$.
\end{enumerate}

\subsection{Factoring polynomials and roots-finding}
\subsubsection{Exhaustive search algorithms}
The problem of finding roots of a polynomial $\Lambda(x)=1+ \lambda_1 x+\cdots + \lambda_tx^t$
could be solved by an exhaustive search in time $O(tp^m)$. Alternatively, one 
may use Fast Fourier Transform that we have discussed in the preceding 
sections to find roots of $\Lambda(x)$ using at most $m^2p^m\log^2(p)$ steps.
Furthermore, one may also use Chien's search to find roots of $\Lambda(x)$.
Chien's search is based on the following observation.
$$\begin{array}{lll}
 \Lambda(\alpha^i)&=&1+ \lambda_1 \alpha^i+\cdots + \lambda_t(\alpha^i)^t\\
&=&1+ \lambda_{1,i}+\cdots + \lambda_{t,i}\\
 \Lambda(\alpha^{i+1})&=&1+ \lambda_1 \alpha^{i+1}+\cdots + \lambda_t(\alpha^{i+1})^t\\
&=&1+ \lambda_{1,i}\alpha+\cdots + \lambda_{t,i}\alpha^t\\
&=&1+ \lambda_{1,i+1}+\cdots + \lambda_{t,i+1}\\
\end{array}$$
Thus, it is sufficient to compute the set $\{ \lambda_{j,i}: i=1, \cdots, q-1; j=1, \cdots, t\}$
with $ \lambda_{j,i+1}= \lambda_{j,i}\alpha^j$. Chien's algorithm can be used 
to improve the performance of RLCE encryption schemes when 64-bits $\oplus$ is used for 
parallel field additions. For non-64 bits CPUs, Chien does not provide advantage over 
exhaustive search algorithms. For the security parameters $128$, Chien's search has better performance
than FFT based search. For the security parameters 192 and 256, FFT based search has better 
performance than Chien's search.

\subsubsection{Berlekamp Trace Algorithm}
Berlekamp Trace Algorithm (BTA) can find the roots of a degree $t$ polynomial in
time $O(mt^2)$. A polynomial $f(x)=f_0+f_1x+\cdots+f_tx^t$ has no repeated roots
if $\gcd(f(x), f'(x))=1$. Without loss of generality, we may assume that $f(x)$ has no repeated 
roots. For each $x\in GF(p^m)$, the trace of $x$ is defined as 
$$\mbox{Tr}(x)=\sum_{i=0}^{m-1} x^{p^i}.$$
We recall that if we consider $GF(p^m)$ as a $m$-dimensional vector space 
over $GF(p)$, then a trace function is linear. That is,
$\mbox{Tr}(ax+by)=\mbox{Tr}(ax)+\mbox{Tr}(bx)$ for $a,b\in GF(p)$ and 
$x,y\in GF(p^m)$. Furthermore, we have $\mbox{Tr}(x^p)=\mbox{Tr}(x)$ for $x\in GF(p^m)$ 
and $\mbox{Tr}(a)=ma$ for $a\in GF(p)$.
It is known that in $GF(p^m)$, we have 
\begin{equation}
\label{traceEq}x^{p^m}-x=\prod_{s\in GF(p)}\left(\mbox{Tr}(x)-s\right).
\end{equation}
Let $\alpha$ be the root of a primitive polynomial of degree $m$ over $GF(p)$. Then 
$(1, \alpha, \cdots, \alpha^{{m-1}})$ is a polynomial basis for $GF(p^m)$ over $GF(p)$ and 
$(\alpha, \cdots, \alpha^{p^{m-1}})$ is a normal basis for $GF(p^m)$ over $GF(p)$.
Substituting $\alpha^ix$ for $x$ in equation (\ref{traceEq}), we get
$$(\alpha^i)^{p^m}x^{p^m}-\alpha^ix=\prod_{s\in GF(p)}\left(\mbox{Tr}(\alpha^ix)-s\right).$$
This implies 
$$x^{p^m}-x=\alpha^{-i}\prod_{s\in GF(p)}\left(\mbox{Tr}(\alpha^ix)-s\right).$$
If $f(x)$ is a nonlinear polynomial that splits in $GF(p^m)$, then $f(x)|(x^{p^m}-x)$. Thus we have 
\begin{equation}
\label{enafac}
f(x)=\prod_{s\in GF(p)} \gcd\left(f(x), \mbox{Tr}(\alpha^ix)-s\right).
\end{equation}
By applying equation (\ref{enafac}) with $i=0,1, \cdots, {m-1}$ or $i=1,p, \cdots, p^{m-1}$, we can  factor $f(x)$. 
In order to speed up the computation of $\mbox{Tr}(\alpha^ix)$ modulo $f(x)$, one pre-computes
the residues of $x, x^2, \cdots, x^{p^m}$ modulo $f(x)$. By adding these residues, one
gets the residue of $\mbox{Tr}(x)$. Furthermore, by multiplying these residues with 
$\alpha^i, \alpha^{2i}, \cdots, \alpha^{ip^m}$ respectively, one obtains the residue of $\mbox{Tr}(\alpha^ix)$. 

For RLCE implementation over $GF(2^m)$, the BTA algorithm can be described as follows.

\vskip 5pt

\noindent
{\em Input:} A polynomial $f(x)$ and pre-compute $\mbox{Tr}_i(x)=x^{2^i}\mod f(x)$ for $i=1, \cdots, m$.

\noindent
{\em Output:} A list of roots $(r_0, \cdots, r_{n_f})=\mbox{BTA}(f(x))$.

\noindent
{\em Algorithm:} 
\begin{enumerate}
\item Let $j=0$.
\item If $f(x)=x+\alpha$, return $\alpha$.
\item Use $\mbox{Tr}_i(x)$ to compute $\mbox{Tr}(\alpha^j x)\mod f(x)$.
\item If $j>m$, return $\emptyset$.
\item Let $p(x)=\gcd(\mbox{Tr}(\alpha^j x), f(x))$ and $q(x)=\frac{f(x)}{p(x)}$.
\item Let $j=j+1$ and return $\mbox{BTA}(p(x)) \cup \mbox{BTA}(q(x))$.
\end{enumerate}

BTA algorithm converts one multiplication into several additions. In RLCE scheme, field multiplication 
is done via table look up. Our experiments show that  BTA algorithm is slower than Chien's search 
or exhaustive search algorithms for RLCE encryption scheme. 

\subsubsection{Linearized and affine polynomials}
\label{linearsec}
In the preceding section, we showed how to compute the roots of polynomials using BTA
algorithm. In practice, one factors a polynomial using BTA algorithm until degree four or less.
For polynomials of lower degrees (e.g., lower than 4), one can use affine multiple of polynomials
to find the roots of the polynomial more efficiently (see., e.g., Berlekamp \cite[Chapter 11]{berlekamp1968algebraic}). 
We first note that a linearized polynomial over $GF(p^m)$ is a polynomial of the form
$$g(x) =\displaystyle\sum_{i=0}^ng_ix^{p^i}$$
with $g_i\in GF(p^m)$. Note that for a linearized polynomial $g$, we have 
$g(ax+by)=g(ax)+g(bx)$ for $a,b\in GF(p)$ and 
$x,y\in GF(p^m)$. 
An affine polynomial is a polynomial in the form $a(x)=g(x)+a$ where
$g(x)$ is a linearized polynomial and $a\in GF(p^m)$.
For small degree polynomials, one can convert it to an affine polynomial which is a multiple of the 
given polynomial. The root of the affine polynomial could be found by solving a linear equation
system of $m$ equations. 

The roots of a degree $t$ polynomial $f(x)$ are calculated as follows. At step $i\ge 0$,
one computes a degree $2^{\lceil \log_2 t\rceil+i}$ affine multiple of $f(x)$. 
The roots of the affine polynomial could be found by solving the following linear equation system of order $m$ over $GF(2)$.
If the system has no solution, one moves to step $i+1$.

Let $A(x)=g(x)+c=\displaystyle\sum_{i=0}^ng_ix^{p^i}+c$  be an affine polynomial and 
$\alpha^0, \alpha, \cdots, \alpha^{m-1}$ be a polynomial basis for $GF(2^m)$ over $GF(2)$.
Let $c=c_0\alpha^0+ \cdots+c_{m-1}\alpha^{m-1}$ and 
$x=x_0\alpha^0+\cdots +x_{m-1}\alpha^{m-1}\in GF(2^m)$ be a root for $A(x)$. Then we have 
the following linear equation system:
$$\begin{array}{ll}
A(x)=0 &\iff g(x)=c\\
&\iff g\left(\displaystyle\sum_{i=0}^{m-1}x_i\alpha^i\right)=\displaystyle\sum_{i=0}^{m-1}x_i\cdot g(\alpha^i)=\displaystyle\sum_{i=0}^{m-1}c_i\alpha^i=c\\
&\iff \displaystyle\sum_{i=0}^{m-1}\left(x_i\displaystyle\sum_{j=0}^{n}g_{j}\alpha^{ip^j}\right)=\displaystyle\sum_{i=0}^{m-1}c_i\alpha^i\\
&\iff \displaystyle\sum_{i=0}^{m-1}\left(x_i\displaystyle\sum_{j=0}^{m-1}e_{i,j}\alpha^j\right)=\displaystyle\sum_{i=0}^{m-1}c_i\alpha^i\\
&\iff \displaystyle\sum_{i=0}^{m-1}\left(\alpha^i\displaystyle\sum_{j=0}^{m-1}x_je_{j,i}\right)=\displaystyle\sum_{i=0}^{m-1}c_i\alpha^i\\
\end{array}
$$
That is, $c_i=\displaystyle\sum_{j=0}^{m-1}x_je_{j,i}$ for $i=0, \cdots, m$ where 
$e_j=(e_{j,0}, \cdots, e_{j,m-1})=\displaystyle\sum_{i=0}^{n}g_{i}\alpha^{jp^i}$.  
The linear system could also be  written as:
\begin{equation}
\left(\begin{array}{cccc}
e_{0,0} & e_{1,0} & \cdots & e_{m-1,0}\\
e_{0,1} & e_{1,1} & \cdots & e_{m-1,1}\\
\vdots & \vdots & \ddots & \ldots\\
e_{0,m-1} & e_{1,1} & \cdots & e_{m-1,m-1}\\
\end{array}\right)
\left(\begin{array}{c}
x_0\\
x_1\\
\vdots\\
x_{m-1}
\end{array}\right)=
\left(\begin{array}{c}
c_0\\
c_1\\
\vdots\\
c_{m-1}
\end{array}\right)
\end{equation}
For the affine polynomial $x^2+ax+c$. We consider two cases. For $a=0$, the square root of $c$ could be calculated directly
as $c^{p^{m-1}}$. For $a\not=0$, we substitute $x$ with $x=ay$ and obtain a new polynomial
$y^2+y+\frac{c}{a^2}$. Thus we have $e_j=\alpha^j+\alpha^{2j}$ which could be pre-computed.
For a polynomial $p(x)=x^3+ax^2+bx+c$, it has a degree 4 affine multiple polynomial 
$p_1(x)=(x+a)(x^3+ax^2+bx+c)=x^4+(a^2+b)x^2+(ab_1+c)x+ac$.
For a degree 4 polynomial $p(x)=x^4+ax^3+bx^2+cx+d$, let $x=y+\sqrt{\frac{c}{a}}$.
We obtain $p(y)=y^4+ay^3+(a\sqrt{\frac{c}{a}}+b)y^2+(\frac{cb}{a}+d)$. Next let $z=\frac{1}{y}$.
Then we have the affine polynomial $p(z)=z^4+\frac{a\sqrt{\frac{c}{a}}+b)}{\frac{bc}{a}+d}z^2+\frac{a}{\frac{cb}{a}+d}z+ 
\frac{1}{\frac{cb}{a}+d}$.
For the affine polynomial $x^4+ax^2+bx+c$,
we have $e_j=b\alpha^j +a\alpha^{2j}+\alpha^{4j}$. 
For the affine polynomial $x^8+ax^4+bx^2+dx+c$,
we have $e_j=d\alpha^j +b\alpha^{2j}+a\alpha^{4j}+\alpha^{8j}$. 

As a special case, we consider the roots for quadratic polynomials over the finite fields $GF(2^{10})$ and $GF(2^{11})$. 
For $p(x)=x^2+x+c$ over $GF(2^{m})$ with $c\not=0$, $p(x)$ has a root if and only if $\mbox{Tr}(x)=0$. 
Let $c=c_0+c_1\alpha+\cdots +c_{m-1}\alpha^{m-1}$ and $\mbox{Tr}(x)=0$. Then the roots for $p(x)$ 
are $x=x_0+x_1\alpha+\cdots +x_{m-1}\alpha^{m-1}$ and $x+1$ where
\begin{enumerate}
\item If $m=10$, then
$$\begin{array}{ll}
x_9&=c_3+c_5+c_6+c_9\\
x_8&=c_3+c_5+c_6\\
x_7&=c_0+c_1+c_2+c_4+c_5+c_8+c_9\\
x_6&=c_0+c_5\\
x_5&=c_0\\
x_4&=c_8+c_9\\
x_3&=c_0+c_3\\
x_2&=c_0+c_1+c_2+c_3+c_6+c_9\\
x_1&=c_1+c_3+c_5+c_6+c_9\\
x_0&=0\\
\end{array}$$
\item If $m=11$, then
$$\begin{array}{ll}
x_{10}&=c_5+c_7+c_9+c_{10}\\
x_9&=c_3+c_5+c_6+c_9+c_{10}\\
x_8&=c_3+c_6\\
x_7&=c_1+c_2+c_3+c_4+c_5+c_6+c_8+c_{10}\\
x_6&=c_9+c_{10}\\
x_5&=c_3+c_5+c_6+c_8+c_9+c_{10}\\
x_4&=c_1+c_2+c_3+c_4+c_5+c_8+c_{10}\\
x_3&=c_3+c_4+c_5+c_6+c_8+c_9+c_{10}\\
x_2&=c_2+c_3+c_4+c_5+c_6+c_8+c_{10}\\
x_1&=c_0\\
x_0&=0\\
\end{array}$$
\end{enumerate}

\subsection{Matrix multiplication and inverse: Strassen algorithm}
Strassen algorithm is more efficient than the standard matrix multiplication algorithm. 
Assume that $A$ is a $n_1\times n_2$ matrix, $B$ is a $n_2\times n_3$ matrix, and all $n_1, n_2, n_3$
are even numbers. Then $C=AB$ could be computed by first partition $A,B,C$ as follows
$$A=\left( \begin{array}{cc}
A_{1,1} & A_{1, 2}\\
A_{2,1} & A_{2,2}
\end{array}\right), 
B=\left( \begin{array}{cc}
B_{1,1} & B_{1, 2}\\
B_{2,1} & B_{2,2}
\end{array}\right),
C=\left( \begin{array}{cc}
C_{1,1} & C_{1, 2}\\
C_{2,1} & C_{2,2}
\end{array}\right)
$$
where $A_{i,j}$ are $\frac{n_1}{2}\times \frac{n_2}{2}$ matrices, 
$B_{i,j}$ are $\frac{n_2}{2}\times \frac{n_3}{2}$ matrices,  and 
$B_{i,j}$ are $\frac{n_1}{2}\times \frac{n_3}{2}$ matrices.
Then we compute the following $7$ matrices of appropriate dimensions:
$$\begin{array}{l}
M_1 = (A_{1,1}+A_{2,2})(B_{1,1}+B_{2,2})\\
M_2=(A_{2,1}+A_{2,2}) B_{1,1}\\
M_3=A_{1,1}(B_{1,2}-B_{2,2})\\
M_4=A_{2,2}(B_{2,1}-B_{1,1})\\
M_5=(A_{1,1}+A_{1,2})B_{2,2}\\
M_6=(A_{2,1}-A_{1,1})(B_{1,1}+B_{1,2})\\
M_7=(A_{1,2}-A_{2,2})(B_{2,1}+B_{2,2})
\end{array}$$
Next the $C_{i,j}$ can be computed as follows:
$$\begin{array}{l}
C_{1,1} = M_1+M_4-M_5+M_7\\
C_{1,2} = M_3+M_5\\
C_{2,1} = M_2+M_4\\
C_{2,2} = M_1-M_2+M_3+M_6\\
\end{array}$$
The process can be carried out recursively until $A$ and $B$ are small enough (e.g., of dimension around 30)
to use standard matrix multiplication algorithms. Note that if the numbers of rows or columns are odd, we can 
add zero rows or columns to the matrix to make these numbers even.
Please note that in Strassen's original paper, the performance is analyzed for square matrices of dimension 
$u2^v$ where $v$ is the recursive steps  and $u$ is the matrix dimension to stop the recursive process.
For a matrix of dimension $n$, Strassen recommend $n\le u2^v$. Our experiments show that Strassen matrix
multiplication could be used to speed up RLCE encryption scheme for several security parameters.

For matrix inversion, let
$$A=\left( \begin{array}{cc}
A_{1,1} & A_{1, 2}\\
A_{2,1} & A_{2,2}
\end{array}\right), 
A^{-1}=\left( \begin{array}{ll}
C_{1,1} & C_{1, 2}\\
C_{2,1} & C_{2,2}
\end{array}\right)$$
Then we compute
$$\begin{array}{l}
M_1=A_{1,1}^{-1}\\
M_2=A_{2,1}M_1\\
M_3=M_1 A_{1,2}\\
M_4=A_{2,1} M_3\\
M_5=M_4-A_{2,2}\\
M_6=M_5^{-1}\\
C_{1,2} = M_3M_6\\
C_{2,1} = M_6M_2\\
M_7=M_3C_{2,1}\\
C_{1,1} = M_1-M_7\\
C_{2,2} = -M_6\\
\end{array}$$
Similarly, for matrices with odd dimensions, we can add zero rows/columns and identity matrices in the lower right corner
to carry out the computation recursively. 

Strassen inversion algorithm generally has better performance than Gauss elimination based algorithm.
However, it has high incorrect abortion rate. Thus it is not useful for RLCE encrypiton schemes. 
For example, Strassen inversion algorithm will abort on the following matrix over $GF(2^{10})$ 
though its inverse does exist. The following matrix is a common matrix for which the matrix inverse is 
needed in RLCE implementation.
$$\left(\begin{array}{ccccccccccc}
0 &313 &0 &626 &252 &266 &62 &841 &0 &506 &0 \\
0 &0 &0 &636 &389 &357 &852 &638 &0 &869 &0 \\
0 &0 &701 &656 &635 &143 &130 &392 &0 &278 &0 \\
0 &0 &711 &433 &1020 &841 &46 &185 &1000 &369 &0 \\
0 &0 &813 &692&219 &657&579 &0 &13 &777&0 \\
0 &0 &350 &923&632 &270 &950 &0 &228 &105 &0 \\
0 &0 &105 &445 &0 &954 &916 &0 &809 &268 &0 \\
0 &0 &963 &217 &0 &619&903 &0 &566&442 &0 \\
0 &0 &0 &455 &0 &815 &219 &0 &708&242&0 \\
129 &0 &0 &334 &0 &702 &481 &0 &0 &614&0 \\
769 &0 &0 &4&0 &729 &955 &0 &0 &545 &433 \\
\end{array}\right)
$$

Note that in order to avoid the incorrect abortion in Strassen inversion algorithm, one may use the Bunch-Hopcroft 
\cite{bunch1974triangular} triangular factorization approach LUP combined with Strassen inversion algorithm. Since 
the LUP factorization requires additional steps for factorization, it will not improve the performance for
RLCE encryption schemes and we did not implement it. Alternatively, one may use the 
Method of Four Russians for Inversion (M4RI) \cite{bard2006accelerating} to speed up the matrix inversion
process. Our analysis shows that the M4RI performance gain for RLCE encryption scheme 
is marginal. Thus we did not implement it either.

\subsection{Vector matrix multiplication: Winograd algorithm}
Winograd's algorithm can be used to reduce the number of multiplication operations in vector matrix multiplication 
by 50\%. Note that this approach could also be used for matrix multiplication. The algorithm is based on the following
algorithm for inner product computation of two vectors $x=(x_0, \cdots, x_{n-1})$ and $y=(y_0, \cdots, y_{n-1})$. 
We first compute 
$$\bar{x}=\sum_{j=0}^{\left\lfloor\frac{n}{2}-1 \right\rfloor} x_{2j}x_{2j+1}\quad\quad\mbox{and}\quad\quad
\bar{y}=\sum_{j=0}^{\left\lfloor\frac{n}{2}-1 \right\rfloor} y_{2j}y_{2j+1}
$$
Then the inner product $x\cdot y$ is given by
$$x\cdot y=
\left\{\begin{array}{ll}
\displaystyle\sum_{j=0}^{\left\lfloor\frac{n}{2}-1 \right\rfloor} (x_{2j}+y_{2j+1})(x_{2j+1}+y_{2j}) -\bar{x}-\bar{y} & n \mbox{ is even} \\
\displaystyle\sum_{j=0}^{\left\lfloor\frac{n}{2}-1 \right\rfloor} (x_{2j}+y_{2j+1})(x_{2j+1}+y_{2j}) -\bar{x}-\bar{y}+x_{n-1}y_{n-1} & n \mbox{ is odd} \\
\end{array}\right.$$

The Winograd algorithm converts each field multiplication into several field additions. Our experiments show
that Winograd algorithm is extremely slow for RLCE encryption implementations when table look up is used for 
field multiplication.

\subsection{Experimental results}
We have implemented these algorithms that we have discussed in the preceding sections. Table \ref{rootper} gives experimental results
on finding roots of error loator polynomials in RLCE schemes. The implementation was run on a MacBook Pro
with masOS Sierra version 10.12.5 with 2.9GHz Intel Core i7 Processor. The reported time is the required milliseconds 
for finding roots of a degree $t$ polynomial over $GF(2^{10})$ (an average of 10,000 trials). These results 
show that generally Chien's search is the best choice.

\begin{table}[h]
\caption{Milliseconds for finding roots of a degree $t$ error locator polynomial over $GF(2^{10})$}
\label{rootper}
\begin{center}
$\begin{array}{|c|c|c|c|c|}\hline
t & \mbox{FFT}& \mbox{Chien Search} & \mbox{Exhaustive search} & \mbox{BTA} \\ \hline
78&.4781572&.2871678&.7360182&1.1814685\\ \hline
80&.5021798&.2864403&.7506306&1.2784691\\ \hline
114&.6632026&.4155929&1.0445943&1.9991356 \\ \hline
118&.6892365&.4280331&1.0773125&2.1493591\\ \hline
230&1.3742336&.8323220&2.0717924&5.7388549\\ \hline
280&1.7690640&1.0194170&2.4806118&8.3730290\\ \hline
\end{array}$
\end{center}
\end{table}

On the other hand, for small degree polynomials, Chien's search might be the best choice. 
Table \ref{rootsmallper} gives experimental results
on finding roots of small degree polynomials. These polynomial degrees are the common 
degrees for polynomials in list-decoding
based RLCE schemes. The implementation was run on a MacBook Pro
with masOS Sierra version 10.12.5 with 2.9GHz Intel Core i7 Processor. The reported time is the required milliseconds 
for finding roots of a degree $t$ polynomial over $GF(2^{10})$ (an average of 10,000 trials). These results 
show that for degree 4 or less, the linearized and affine polynomial based BTA is the best choice. For degrees above 4, Chien's search
is the best choice.

\begin{table}[h]
\caption{Milliseconds for finding roots of a small degree $t$ polynomial over $GF(2^{10})$}
\label{rootsmallper}
\begin{center}
$\begin{array}{|c|c|c|c|c|}\hline
t & \mbox{Chien Search} & \mbox{BTA}& \mbox{FFT}& \mbox{Exhaustive search}  \\ \hline
4&.0197496&.0009202&.1117984&.1175816 \\ \hline
6&.0261202&.0537054&.1174620&.1252327\\ \hline
8&.0330730&.1215397&.1402607&.1419983 \\ \hline
10&.0418521&.1288605&.1417330&.1605130\\ \hline
14&.0537797&.1780427&.1481447&.1908748\\ \hline
18&.0669920&.2288600&.1805597&.2228205\\ \hline
\end{array}$
\end{center}
\end{table}

Table \ref{polymulper} gives experimental results for RLCE polynomial multiplications. 
The implementation was run on a MacBook Pro
with masOS Sierra version 10.12.5 with 2.9GHz Intel Core i7 Processor. 
The reported time is the required milliseconds  for multiplying a degree $t$ polynomial with 
a  degree $2t$ polynomial over $GF(2^{10})$ (an average of 10,000 trials).
From the experiment, it shows that Karatsuba's polynomial algorithm only
outperforms standard polynomial algorithm for polynomisl degrees 
above degree 115. It is noted that in standard test, 
Karatsuba's polynomial algorithm outperforms standard polynomial algorithm for polynomial degrees 
above degree 35 already.

\begin{table}[h]
\caption{Milliseconds  for multiplying a pair of degree $t$ and $2t$ polynomials over $GF(2^{10})$}
\label{polymulper}
\begin{center}
$\begin{array}{|c|c|c|c|}\hline
t & \mbox{Karatsuba}& \mbox{Standard Algorithm}  & \mbox{FFT} \\ \hline
78& .0470269&.0374369&1.4651561\\ \hline
80& .0546122&.0423766&1.4891211\\ \hline
114&.0794242&.0775524&2/4723263\\ \hline
118&.0811117&.0833309&2.5360034\\ \hline
230&.2371405&.3117507&6.3380415\\ \hline
280&.3444224&.4547458&7.8866734\\ \hline
\end{array}$
\end{center}
\end{table}

Table \ref{matmulper} gives experimental results for RLCE related matrix multiplications. 
The implementation was run on a MacBook Pro
with masOS Sierra version 10.12.5 with 2.9GHz Intel Core i7 Processor. 
The reported time is the required seconds for multiplying two $n\times n$ matrices 
(or invert an $n\times n$ matrix) over $GF(2^{10})$ (an average of 100 trials)..

\begin{table}[h]
\caption{Seconds for multiplying a pairs of (inverting a) $n\times n$ matrices over $GF(2^{10})$}
\label{matmulper}
\begin{center}
$\begin{array}{|c|c|c|c|c|c|}\hline
n & \mbox{Strassen Mul.} & \mbox{Standard Mul.}& \mbox{Winograd Mul.} &  \mbox{Gauss Elimination Inv}& \mbox{Strassen Inv.}\\ \hline
376&.17881616&.15684892&.57614453&.23071715&.22307581\\ \hline
470 &.42498317&.30317405&1.12305698&.44601063&.53218560\\ \hline
618 &.77971244&.65356388&2.68176523&.97155253&.98632941\\ \hline
700&1.01458090&.94067030&3.77942598&1.41453963&1.30181261\\ \hline
764&1.20244299&1.21845951&4.88860081&1.82576160&1.55965069\\ \hline
800&1.36761960&1.605249880&6.27596202&2.14227823&1.80930063\\ \hline
\end{array}$
\end{center}
\end{table}

\section{Reed-Solomon codes}
\label{rssec}
\subsection{The original approach}
Let $k<n< q$ and $a_0, \cdots, a_{n-1}$ be distinct elements from $GF(q)$. The Reed-Solomon
code is defined as 
$${\cal C}=\left\{(m(a_0), \cdots, m(a_{n-1})): m(x) \mbox{ is a polynomial over } GF(q) \mbox{ of degree }<k\right\}.$$
There are two ways to encode $k$-element messages within Reed-Solomon codes.
In the original approach, the coefficients of the polynomial 
$m(x)=m_0+m_1x+\cdots+m_{k-1}x^{k-1}$ is considered as the message 
symbols. That is, the generator matrix $G$ is defined as 
$$G=\left( 
\begin{array}{ccc}
1 & \cdots & 1\\
a_0 &\cdots & a_{n-1}\\
\vdots & \ddots & \vdots\\
a_0^{k-1} &\cdots & a_{n-1}^{k-1}
\end{array}
\right)$$
and the the codeword for the message symbols $(m_0, \cdots, m_{k-1})$ is 
$(m_0, \cdots, m_{k-1})G$. 

Let $\alpha$ be a primitive element of 
$GF(q)$ and $a_i=\alpha^i$.  Then it is observed
that Reed-Solomon code is cyclic when $n=q-1$. 
For each $j>0$,  let 
${\bf m}=(m_0, \cdots, m_{k-1})$ and 
${\bf m}'=(m_0\alpha^0, m_1\alpha^1, \cdots, m_{k-1}\alpha^{k-1})$.  
Then  $m'(\alpha^i)=m_0\alpha^0+m_1\alpha^1\alpha^i+\cdots+m_{k-1}\alpha^{k-1}\alpha^{i(k-1)}
=m(\alpha^{i+1})$. That is,
${\bf m}'$ is encoded as 
$$\left(m'(\alpha^0), \cdots, m'(\alpha^{n-1})\right)=\left(m(\alpha), \cdots,
m(\alpha^{n-1}), m(\alpha^{0})\right)$$
which is a cyclic shift of the codeword for ${\bf m}$.

Instead of using  coefficients to encode messages, one may use
$m(a_0), \cdots, m(a_{k-1})$ to encode the message symbols. This
is a systematic encoding approach and one can encode a message vector using
Lagrange interpolation.

\subsection{The BCH approach}
\label{bchsec}
We first give a definition for the $t$-error-correcting BCH codes of distance $\delta$. 
Let $1\le \delta<n=q-1$ and let $g(x)$ be a polynomial over $GF(q)$ such that 
$g(\alpha^b)=g(\alpha^{b+1})=\cdots =g(\alpha^{b+\delta-2})=0$
where  $\alpha$ is a primitive $n$-th
root of unity (note that it is not required to have $\alpha\in GF(q)$). 
It is straightforward to check that $g(x)$ is a factor of $x^n-1$. 
For $w=n-\deg(g)-1$, a message polynomial $m(x)=m_0+m_1x+\cdots +m_wx^{w}$
over $GF(q)$ is encoded as a degree $n-1$ polynomial $c(x)=m(x)g(x)$. 
A BCH codes with $b=1$ is called a narrow-sense BCH code.
A BCH code with $n=q^m-1$ is called a primitive BCH code where $m$ 
is the multiplicative order of $q$ modulo $n$. That is, $m$ is the 
least integer so that $\alpha\in GF(q^m)$.

A BCH code with $n=q-1$ and $\alpha\in GF(q)$
is called a Reed-Solomon code. Specifically, let $1\le k<n=q-1$ and let 
$g(x)=(x-\alpha^b)(x-\alpha^{b+1})\cdots (x-\alpha^{b+n-k-1})=g_0+g_1x+\cdots+g_{n-k}x^{n-k}$
be a polynomial over $GF(q)$. Then a message polynomial $m(x)=m_0+m_1x+\cdots +m_{k-1}x^{k-1}$
is encoded as a degree $n-1$ polynomial $c(x)=m(x)g(x)$. 
In other words, the Reed-Solomon code is the cyclic code generated by the polynomial
$g(x)$. The generator matrix for this definition
is as follows:
$$G=\left(\begin{array}{ccccccc}
g_0 & g_1 &\cdots &g_{n-k} & 0 & \cdots & 0\\
0&g_0 &\cdots &g_{n-k-1} & g_{n-k}  & \cdots & 0\\
\vdots&\vdots &\ddots &\vdots & \vdots  & \ddots & \vdots\\
0&0 &\cdots &g_{n-2k+1} & g_{n-2k+2}  & \cdots & g_{n-k} \\
\end{array}\right)=\left(\begin{array}{c}
g(x)\\
xg(x)\\
\vdots\\
x^{k-1}g(x)
\end{array}\right)$$

For BCH systematic encoding, we first choose the coefficients
of the $k$ largest monomials of $c(x)$ as the message symbols. Then we 
set the remaining coefficients of $c(x)$ in such a way that $g(x)$ divides $c(x)$.
Specifically, let 
$c_r(x)=m(x)\cdot x^{n-k} \mod g(x)$
which has degree $n-k-1$. Then $c(x)=m(x)\cdot x^{n-k} -c_r(x)$ is a systematic 
encoding of $m(x)$. The code polynomial $c(x)$ can be computed by 
simulating a LFSR with degree $n-k$
where the feedback tape contains the coefficients of $g(x)$.

\subsection{The equivalence}
The equivalence of the two definitions for Reed-Solomon code could be 
established using the relationship between FFT and IFFT. 
For each Reed-Solomon codeword $f(x)$ in the BCH approach, 
it is a multiple of the generating polynomial
$g(x)=\displaystyle\prod_{j=1}^{n-k}\left(x-\alpha^j\right)$. Let $F(x)$ be defined as in (\ref{invFFT}). 
Since $f(\alpha^j)=0$ for $1\le j\le n-k$, $F(x)$ has degree at most $k-1$.
By the identity (\ref{fdfdfghifft}),  we have 
$$\mbox{FFT}(F(x))=\left(F(\alpha^0), \cdots, F(\alpha^{n-1})\right)=n\cdot f(x).$$
Thus $f(x)$ is also a Reed-Solomon codeword in the original approach.

For each Reed-Solomon codeword $(a_0, \cdots, a_{n-1})$ in the original approach, it is 
an evaluation of a polynomials $F(x)$ of degree at most $k-1$ on 
$\alpha^0, \cdots, \alpha^{n-1}$. Let $f(x)$ be the function satisfying the identity
(\ref{invFFT}) obtained by interpolation. Then $f(x)=\mbox{FFT}\left(\frac{F(x)}{n}\right)$,
$(a_0, \cdots, a_{n-1})$ is the coefficients of $n\cdot f(x)$,
and $f(\alpha^j)=0$ for $j=1,\cdots, n-k$. Thus $f(x)$ is a multiple
of the generating polynomial $g(x)$.

\subsection{Generalized Reed-Solomon codes}
For an $[n,k]$ generator matrix $G$ for a Reed-Solomon code, we can select
$n$ random elements $v_0, \cdots, v_{n-1}\in GF(q)$ and define a new generator matrix
$$G(v_0,\cdots, v_{n-1})=
G\left(\begin{array}{cccc}
v_0 & 0 & \cdots& 0\\
0 & v_1 & \cdots& 0\\
\vdots & \vdots & \ddots& \vdots\\
0 & 0 & \cdots& v_{n-1}\\
 \end{array}\right)=G\cdot \mbox{diag}{(v_0,\cdots, v_{n-1})}.$$
The code generated by $G(v_0,\cdots, v_{n-1})$ is called a generalized Reed-Solomon code.
For a generalized Reed-Solomon codeword ${\bf c}$, it is straightforward 
that ${\bf c}\cdot \mbox{diag}{\left(v^{-1}_0,\cdots, v^{-1}_{n-1}\right)}$ is a Reed-Solomon codeword.
Thus the problem of decoding generalized Reed-Solomon codes could be 
easily reduced to the problem of decoding Reed-Solomon codes.

\section{Decoding Reed-Solomon code}
\label{decoder}
\subsection{Peterson-Gorenstein-Zierler decoder}
This sections describes Peterson-Gorenstein-Zierler decoder which 
has computational complexity $O(n^3)$.
Assume that Reed-Solomon code is based on BCH approach   
and the received polynomial is
$$r(x)=c(x)+e(x)=r_0+r_1x+\cdots+r_{n-1}x^{n-1}.$$
We first calculate the syndromes $S_{\!j}=r(\alpha^j)$ for $j=1,\cdots, n-k$.
 $$\begin{array}{lll}
 S_j &=&r_0+ r_1 \alpha^j+\cdots + r_{n-1}(\alpha^j)^{n-1}\\
&=&r_0+ r_{1,j}+\cdots + r_{n-1,j}\\
S_{j+1}&=&r_0+ r_1 \alpha^{j+1}+\cdots + r_{n-1}(\alpha^{j+1})^{n-1}\\
&=&r_0+ r_{1,j}\alpha+\cdots + r_{n-1,j}\alpha^{n-1}\\
&=&r_0+ r_{1,j+1}+\cdots + r_{n-1,j+1}\\
\end{array}$$
From the above equations, it is sufficient to compute the set $\{r_{i,j}: i=1, \cdots, n-1; j=1, \cdots, n-k\}$
with $r_{i,j+1}= r_{i,j}\alpha^i$ and then add them together to get the syndromes.

Let the numbers $0\le p_1, \cdots, p_{t}\le n-1$ be error positions  and 
$e_{p_i}$ be error magnitudes (values). Then 
$$e(x)=\sum_{i=1}^{t}e_{p_i}x^{p_i}.$$ 
For convenience,
we will use $X_i=\alpha^{p_i}$ to denote error locations
and $Y_i=e_{p_i}$ to denote error magnitudes.
It should be noted that for the syndromes $S_{\!j}$
for $j=1,\cdots, n-k$, we have 
$$S_{\!j}=r(\alpha^j)=c(\alpha^j)+e(\alpha^j)=e(\alpha^j)
=\sum_{i=1}^{t}e_{p_i}(\alpha^j)^{p_i}=\sum_{i=1}^{t}Y_iX_i^j.$$
That is, we have 
\begin{equation}
\label{equatin}
\left(\begin{array}{cccc}
X_1^1 & X_2^1 & \cdots & X_{t}^1\\
X_1^2 & X_2^2 & \cdots & X_{t}^2\\
\vdots & \vdots  & \ddots & \vdots \\
X_1^{n-k} & X_2^{n-k} & \cdots & X_{t}^{n-k}\\
\end{array}\right)
\left(\begin{array}{c}
Y_1\\
Y_2\\
\vdots\\
Y_{t}
\end{array}
 \right)
=
\left( \begin{array}{c}
S_1\\
S_2\\
\vdots\\
S_{n-k}
\end{array}\right)
\end{equation}
Thus we obtained $n-k$ equations with $n-k$ unknowns: 
$X_1, \cdots, X_t, Y_1, \cdots, Y_t$.
The error locator polynomial is defined as
\begin{equation}
\label{elp}
 \Lambda(x)=\prod_{i=1}^t(1-X_ix)=1+ \lambda_1 x+\cdots + \lambda_tx^t.
\end{equation}
Then we have 
\begin{equation}
\label{fafd}
 \Lambda(X_i^{-1})=1+ \lambda_1 X_i^{-1}+\cdots + \lambda_tX_i^{-t}=0\quad\quad (i=1, \cdots, t)
\end{equation}
Multiply both sides of (\ref{fafd}) by $Y_iX_i^{j+t}$, we get 
\begin{equation}
\label{mafdaf}
Y_iX_i^{j+t} \Lambda(X_i^{-1})=Y_iX_i^{j+t}+ \lambda_1 Y_iX_i^{j+t-1}+\cdots +  \lambda_tY_iX_i^j=0
\end{equation}
For $i=1, \cdots, t$, add equations (\ref{mafdaf}) together, we obtain
\begin{equation}
\label{gtrwwtr}
\sum_{i=1}^t (Y_iX_i^{j+t}) + \lambda_1 \sum_{i=1}^t (Y_iX_i^{j+t-1}) +\cdots + \lambda_t\sum_{i=1}^t (Y_iX_i^{j})=0
\end{equation}
Combing (\ref{equatin}) and (\ref{gtrwwtr}), we obtain 
\begin{equation}
\label{teqar}
S_j \lambda_t+S_{j+1} \lambda_{t-1}+\cdots +S_{j+t-1} \lambda_1+S_{j+t}=0
\quad\quad (j=1,\cdots, t)
\end{equation}
which yields the following linear equation system:
\begin{equation}
\label{elpeq}
\left(\begin{array}{cccc}
S_1 & S_2 & \cdots & S_{t}\\
S_2 & S_3 & \cdots & S_{t+1}\\
\vdots & \vdots  & \ddots & \vdots \\
S_t & S_{t+1} & \cdots & S_{2t-1}\\
\end{array}\right)
\left(\begin{array}{c}
 \lambda_t\\
 \lambda_{t-1}\\
\vdots\\
 \lambda_1
\end{array}
 \right)
=
\left( \begin{array}{c}
-S_{t+1}\\
-S_{t+2}\\
\vdots\\
-S_{2t}
\end{array}\right)
\end{equation}
Since the number of errors is unknown, Peterson-Gorenstein-Zierler tries 
various $t$ from the maximum $\frac{n-k}{2}$ to solve the equation system 
(\ref{elpeq}). After the error locator polynomial $ \Lambda(x)$ is identified, one
can use exhaustive search algorithm, Chien's search algorithm, BTA algorithms, 
or other root-finding algorithms
to find the roots of $\Lambda(x)$. After the error locations 
are identified, one can use Forney's algorithm to determined the error values.
With $e(x)$ in hand, one subtracts $e(x)$ from $r(x)$ to obtain $c(x)$.

\noindent
{\bf Computational complexity:} Assume that $(\alpha^j)^i$ for $i=0, \cdots, n-1$ 
and $j=0, \cdots, n-k$ have been pre-computed in a table. Then it takes 
$2(n-1)(n-k)$ field operations to compute the values of $S_1, \cdots, S_{n-k}$.
After $S_i$ are computed, it takes $O(t^3)$ field operations (for Gaussian eliminations) to 
solve the equation (\ref{elpeq}) for each chosen $t$.

\subsubsection{Forney's algorithm}
For Forney's algorithm, we define the error evaluator polynomial (note that $n-k\ge 2t$)
\begin{equation}
\Omega(x) = \Lambda(x)+\sum_{i=1}^t X_iY_ix\prod_{j=1, j\not= i}^t(1-X_jx)
\end{equation}
and the  syndrome polynomial  
$$S(x)=S_1x+S_2x^2+\cdots S_{2t}x^{2t}.$$
Note that 
\begin{equation}
\begin{array}{lll}
S(x) \Lambda(x)&=& \left(\displaystyle\sum_{l=1}^{2t}\displaystyle\sum_{i=1}^tY_iX_i^lx^l\right)\displaystyle\prod_{j=1}^t(1-X_jx)\quad \mod x^{2t+1}\\
&=&\displaystyle\sum_{i=1}^tY_i\displaystyle\sum_{l=1}^{2t}(X_ix)^l \displaystyle\prod_{j=1}^t(1-X_jx)\quad \mod x^{2t+1}\\
&=&\displaystyle\sum_{i=1}^tY_i(1-X_ix)\displaystyle\sum_{l=1}^{2t}(X_lx)^i\displaystyle\prod_{j=1, j\not=i}^t(1-X_jx)\quad \mod x^{2t+1}\\
\end{array}
\end{equation}
Using the fact that $(1-x^{2t+1})=(1-x)(1+x+\cdots+x^{2t})$, we have 
$$(1-X_ix)\sum_{l=1}^{2t}(X_ix)^l=X_ix-(X_ix)^{2t+1}=X_ix\mod x^{2t+1}.$$
Thus 
$$S(x) \Lambda(x)=\sum_{i=1}^tY_iX_ix\prod_{j=1, j\not=i}^t(1-X_jx)\quad \mod x^{2t+1}.$$
This gives us the key equation
\begin{equation}
\label{keyequation}
\Omega(x)=(1+S(x)) \Lambda(x)\quad\mod x^{2t+1}.
\end{equation}

\noindent
{\bf Note:} In some literature, syndrome polynomial is defined as 
$S(x)=S_1+S_2x+···+S_{2t}x^{2t-1}$.
In this case, the key equation becomes 
\begin{equation}
\label{keyequationvar}
\Omega(x)=S(x) \Lambda(x)\quad\mod x^{2t}.
\end{equation}

Let $ \Lambda'(x)=-\displaystyle\sum_{i=1}^tX_i\displaystyle\prod_{j\not=i}(1-X_jx)=\displaystyle\sum_{i=1}^ti \lambda_ix^{i-1}$.
Then we have $ \Lambda'(X_l^{-1})=-X_l\displaystyle\prod_{j\not=l}(1-X_jX_l^{-1})$. 
By substituting $X_l^{-1}$ into $\Omega(x)$, we get
$$\Omega(X_l^{-1})=\sum_{i=1}^t X_iY_iX_l^{-1}\prod_{j=1, j\not= i}^t(1-X_jX_l^{-1})
=Y_l\prod_{j=1, j\not= l}^t(1-X_jX_l^{-1})=-Y_lX_l^{-1} \Lambda'(X_l^{-1})$$
This shows that 
$$e_{p_l}=Y_l=-\frac{X_l\cdot \Omega(X_l^{-1})}{ \Lambda'(X_l^{-1})}.$$

\noindent
{\bf Computational complexity:} Assume that $(\alpha^j)^i$ for $i=0, \cdots, n-1$ 
and $j=0, \cdots, n-k$ have been pre-computed in a table. Furthermore, assume 
that both $\Lambda(x)$ and $S(x)$ have been calculated already. Then 
it takes $O(n^2)$ field operations to calculate $\Omega(x)$. 
After both $\Omega(x)$ and $\Lambda(x)$ are calculated, it takes
$O(n)$ field operations to calculate each $e_{p_l}$. As a summary, 
assuming that $S(x)$ and $\Lambda(x)$  are known, it takes 
$O(n^2)$ field operations to calculate all error values.

\subsection{Berlekamp-Massey decoder}
In this section we discuss Berlekamp-Massey decoder \cite{massey1969shift} which
has computational complexity $O(n^2)$. Note that there
exists an implementation using Fast Fourier Transform that runs in time $O(n \log n)$.
Berlekamp-Massey algorithm is an alternative approach to find the minimal 
degree $t$ and the error locator polynomial $ \Lambda(x)=1+ \lambda_1x\cdots+ \lambda_tx^t$
such that all equations in (\ref{teqar}) hold. The equations in (\ref{teqar})
define a general linear feedback shift register (LFSR) with initial state $S_1, \cdots, S_t$.
Thus the problem of finding the error locator polynomial $\Lambda(x)$
is equivalent to calculating the linear complexity (alternatively, the 
connection polynomial of the minimal length LFSR) of the sequence  $S_1, \cdots, S_{2t}$. 
The Berlekamp-Massey algorithm constructs an LFSR that produces the entire sequence 
$S_1, \cdots, S_{2t}$ by successively modifying an existing LFSR to produce 
increasingly longer sequences. The algorithm starts with 
an LFSR that produces $S_1$ and then checks whether this LFSR can produce 
$S_1S_2$. If the answer is yes, then no modification is necessary. Otherwise,
the algorithm revises the LFSR in such a way that it can produce $S_1S_2$.
The algorithm runs in $2t$ iterations
where the $i$th iteration computes the linear complexity and connection
polynomial for the sequence  $S_1, \cdots, S_{i}$.  
The following is the original LFSR 
Synthesis Algorithm from Massey \cite{massey1969shift}. 

\begin{center}
\fbox{\begin{minipage}{25em}
\begin{enumerate}[topsep=-1ex,itemsep=-1ex,partopsep=-1ex,parsep=1ex]
\item  $ \Lambda(x)=1, B(x)=1, u=1, L=0, b=1, i=0$.
\item\label{step2} If $i=2t$, stop. Otherwise, compute
\vskip -15pt
\begin{equation}
\label{dispeq}
d=S_i+\sum_{j=1}^L \lambda_jS_{i-j}
\end{equation}
\vskip -15pt
\item If $d=0$, then $u=u+1$, and go to (\ref{step6}).
\item If $d\not=0$ and $i<2L$, then
\vskip -15pt
$$\begin{array}{l}
 \Lambda(x)= \Lambda(x)-db^{-1}x^u B(x)\\
u=u+1
\end{array}$$
\vskip -15pt
and go to (\ref{step6}).
\item If $d\not=0$ and $i\ge 2L$, then
\vskip -15pt
\begin{equation}
\label{lambdadef}\begin{array}{l}
T(x)= \Lambda(x)\\
 \Lambda(x)= \Lambda(x)-db^{-1}x^u B(x)\\
L=i+1-L\\
B(x)=T(x)\\
b=d\\
u=1
\end{array}
\end{equation}
\vskip -15pt
\item\label{step6} $i=i+1$ and go to step (\ref{step2}).
\end{enumerate}
\end{minipage}
}
\end{center}

\noindent
{\bf Discussion:}
For the sequence  $S_1, \cdots, S_{i}$, we use $L_i=L(S_1, \cdots, S_{i})$ to denote
its linear complexity. We use 
$ \Lambda^{(i)}(x)=1+ \lambda_1^{(i)}x+\lambda_2^{(i)}x^2+\cdots+ \lambda_{L_i}^{(i)}x^{L_i}$ 
to denote the connection polynomial for the sequence $S_1\cdots S_i$ that we have obtained
at iteration $i$.   At iteration $i$, the constructed LFSR can produce the 
sequence $S_1S_2\cdots S_i$. That is,
$$S_j=-\sum_{l=1}^{L_i}\lambda_j^{(i)}S_{j-l}, \quad\quad j=L_{i}+1, \cdots, i$$
Let $i_0$ denote 
the last position where the linear complexity changes during the iteration
and let $d_i$ denote the discrepancy obtained at iteration $i$ using the equation (\ref{dispeq}).
That is,
$$d_i=S_i+\sum_{j=1}^{L_{i-1}} \lambda^{(i-1)}_jS_{i-j}.$$
We show that $ \Lambda^{(i)}(x)=\Lambda^{(i-1)}(x)-d_ib^{-1}x^u B(x)$ 
is the connection polynomial for the sequence 
$S_1, \cdots, S_i$. The case for $d_i=0$ is trivial. Assume that $d_i\not=0$. Then
$B(x)= \Lambda^{(i_0)}(x)$ and $b=d_{i_0+1}$. By the construction in Step 4 and Step 5, we have 
$ \Lambda^{(i)}(x)= \Lambda^{(i-1)}(x)-d_id_{i_0+1}^{-1}x^u  \Lambda^{(i_0)}(x)$. For 
$v=L_i, L_i+1, \cdots, i-1$, we have 
$$\begin{array}{lll}
S_v+\sum_{j=1}^{L_i} \lambda_j^{(i)}S_{v-j} &=& S_v+\sum_{j=1}^{L_{i-1}} \lambda_j^{(i-1)}S_{v-j}
   +d_id_{i_0+1}^{-1}\left(S_{v-i+i_0+1}+\sum_{j=1}^{L_{i_0}} \lambda_j^{(i_0)}S_{v-i+i_0+1-j}   \right)\\
&=&\left\{ \begin{array}{ll}
0 & L_i\le u\le i-1\\
d_i-d_id_{i_0+1}^{-1}d_{i_0+1} & u=i
\end{array}\right.
\end{array}$$

\noindent
{\bf Computational complexity:} As we have mentioned in Section \ref{decoder}, it takes 
$2(n-1)(n-k)$ field operations to calculates the sequence $S_1, \cdots, S_{n-k}$.
In the Berlekamp-Massey decoding process, iteration $i$ requires at most $2(i-1)$ field operations to 
calculate $d_i$ and at most $2(i-1)$ operations to calculate the polynomial $ \Lambda^{(i)}(x)$.
Thus it takes at most $4t(2t-1)$ operations to finish the iteration process. In a summary,
Berlekamp-Massey decoding process requires at most $2(n-1)(n-k)+4t(2t-1)$
field operations.

\subsection{Euclidean decoder}
Assume that the polynomial $S(x)$ is known already.
By the key equation (\ref{keyequation}), 
we have 
$$\Omega(x)=(1+S(x)) \Lambda(x)\mod x^{2t+1}$$
with $\deg(\Omega(x))\le \deg( \Lambda(x))\le t$. 
The generalized Euclidean algorithm
could be used to find a sequence of polynomials 
$R_1(x), \cdots, R_u(x)$ , $Q_1(x), \cdots, Q_u(x)$ such that
$$\begin{array}{l}
x^{2t+1}-Q_1(x)(1+S(x))=R_1(x)\\
1+S(x)-Q_2(x)R_1(x)=R_2(x)\\
\cdots \\
R_{u-2}(x)-Q_{u}(x)R_{u-1}(x)=R_u(x)
\end{array}
$$
where $\deg(1+S(x))>\deg(R_{1}(x))$, $\deg(R_i(x))>\deg(R_{i+1}(x))$ ($i=1, \cdots, u-1$),
$\deg(R_{u-1}(x))\ge t$, and $\deg(R_u(x))<t$. By substituting first $u-1$ identities
into the last identity, we obtain the key equation
$$ \Lambda(x)(1+S(x))-\Gamma(x)x^{2t+1}=\Omega(x)$$
where $R_u(x)=\Omega(x)$.

In case that the syndrome polynomial is defined as 
$S(x)=S_1+S_2x+···+S_{2t}x^{2t-1}$, the Euclidean decoder
will calculate the key equation 
$$ \Lambda(x)S(x)-\Gamma(x)x^{2t}=\Omega(x)$$

\noindent
{\bf Computational complexity:}
As we mentioned in the previous sections, it takes $2(n-1)(n-k)$ field 
operations to calculate the polynomial $S(x)$. After $S(x)$ is obtained,
the above process stops in $u$ steps where $u\le t+1$.
For each identity, it requires at most $O(t)$ steps to obtain the pair 
of polynomials $(R_i, Q_i)$. Thus the total steps required by the Euclidean decoder
is bounded by $O(t^2)$. 

\subsection{Berlekamp-Welch decoder}
In previous sections, we dicussed syndrome-based decoding 
algorithms for Reed-Solomon codes. In this and next sections 
we will discuss syndromeless decoding algorithms that 
do not compute syndromes and do not use the Chien search and Forney’s formula.
We first introduce Berlekamp-Welch decoding algorithm 
which has computational complexity $O(n^3)$. 
Berlekamp-Welch decoding algorithm first 
appeared in the US Patent 4,633,470 (1983). The algorithm is based
on the classical definition of Reed-Solomon codes and can be easily adapted 
to the BCH definition of Reed-Solomon codes. The decoding 
problem for the classical Reed-Solomon codes is described as follows:
We have a polynomial $m(x)$ of degree at most $k-1$ and we received
a polynomial $c(x)$ which is given by its evaluations $(r_0, \cdots, r_{n-1})$ on 
$n$ distinct field elements.  We know that $m(x) = r(x)$ for at least 
$n-t$ points. We want to recover $m(x)$ from $r(x)$ efficiently.  

Berlekamp-Welch decoding algorithm is based on the fundamental vanishing lemma
for polynomials:  If $m(x)$ is a polynomial of degree at most $d$ 
and $m(x)$ vanishes at $d+1$ distinct points, then $m$ is the zero polynomial.
Let the graph of $r(x)$ be the set of $q$ points:
$$\left\{(x,y)\in GF(q): y=r(x)\right\}.$$
Let $R(x,y)=Q(x)-E(x)y$ be a non-zero lowest-degree polynomial that vanishes on 
the graph of $r(x)$. That is, $Q(x)-E(x)r(x)$ is the zero polynomial.
In the following, we first show that $E(x)$ has degree at most $t$ and $Q(x)$ 
has degree at most $k+t-1$. 

Let $x_1, \cdots, x_{t'}$ be the list of all positions that 
$r(x_i)\not=m(x_i)$ for $i=1, \cdots, t'$ where $t'\le t$. Let 
$$E_0(x)=(x-x_1)(x-x_2)\cdots (x-x_{t'}) \mbox{ and }Q_0(x)=m(x)E_0(x).$$
By definition, we have $\mbox{deg}(E_0(x))=t'\le t$ and $\mbox{deg}(Q_0(x))=t'+k-1\le t+k-1$.
Next we show that $Q_0(x)-E_0(x)r(x)$ is the zero polynomial.
For each $x\in GF(q)$, we distinguish two cases. For the first case,
assume that $m(x)=r(x)$. Then $Q_0(x) = m(x)E_0(x)=r(x)E_0(x)$. 
For the second case, assume that $m(x)\not=r(x)$. Then $E_0(x)=0$.
Thus we have $Q_0(x)=m(x)E_0(x) = 0 = r(x)E_0(x)$. 
This shows that there is a polynomial $E(x)$ 
of degree at most $t$ and a polynomial $Q(x)$ of degree at most 
$k+t-1$ such that $R(x,y)= Q(x)-E(x)y$ vanishes on the graph of $r(x)$.

The arguments in the preceding paragraph show that, for the minimal degree polynomial
$R(x,y)=Q(x)-E(x)y$, both $Q(x)$ and $m(x)E(x)$ are polynomials of degree 
at most $k+t-1$. Thus $Q(x)-m(x)E(x)$ has degree at most $k+t-1$.
For each $x$ such that $m(x)-r(x)=0$, we have $Q(x)-m(x)E(x)=0$.  
Since $m(x)-r(x)$ vanishes on at least $n-t$ positions
and $n-t>k+t-1$, the polynomial $R(x, m(x))=Q(x)-m(x)E(x)$ 
must be the zero polynomial.

The equation $Q(x)-E(x)r(x)=0$ is called the key equation for the decoding algorithm.
The arguments in the preceding paragraphs show that for any 
solutions $Q(x)$ of degree at most $k+t-1$ and $E(x)$ of degree at most $t$,
$Q(x)-m(x)E(x)$ is the zero polynomial. That is, $m(x)=\frac{Q(x)}{E(x)}$. 
This implies that, after solving the key equation, we can calculate the 
message polynomial $m(x)$.
Let $(m(a_0), \cdots, m(a_{n-1}))$ be the transmitted code and 
$(r_0, \cdots, r_{n-1})$ be the received vector.  Define two polynomials
with unknown coefficients:
$$\begin{array}{l}
Q(x)=u_0+u_1x+\cdots + u_{k+t-1}x^{k+t-1}\\
E(x)=v_0+v_1x+\cdots +v_tx^t
\end{array}$$
Using the identities
$$Q(a_i)=r_i\cdot E(a_i)\quad\quad (i=0, \cdots, n-1)$$
to build a linear equation system of $n$ equations in $n+1$
unknowns $u_0, \cdots, u_{k+t-1}, v_0, \cdots, v_t$.
Find a non-zero solution of this equation system and 
obtain the polynomial $Q(x)$ and $E(x)$. 
Then $m(x)=\frac{Q(x)}{E(x)}$.

\noindent
{\bf Computational complexity:} 
The Berlekamp-Welch decoding process solves 
an equation system of $n$ equations in $n+1$ unknowns.
Thus the computational complexity is $O(n^3)$.

\subsection{List decoder}

Based on Berlekamp-Welch decoding algorithm, Sudan \cite{sudan1997decoding} designed an algorithm
to decode Reed-Solomon codes by correcting up to $n-1-\left\lfloor\sqrt{2n(k-1)}\right\rfloor\ge \frac{n-k}{2}$ errors.
Guruswami and Sudan \cite{guruswami1998improved} improved Sudan's algorithm
to correct up to $t_{GS}(n,k)=n-1-\left\lfloor\sqrt{n(k-1)}\right\rfloor$ errors. 
List-decoding techniques have been used by authors such as Bernstein, Lange, and Peters \cite{bernstein2008attacking}
to improve the security of McEliece encryption schemes.
In this section, we present Guruswami-Sudan's (GS) algorithm with  K\"{o}tter's iterative interpolation \cite{kotter1996fast} and 
Roth-Ruckenstein's polynomial factorization \cite{roth2000efficient}. 

For a message polynomial $m(x)=m_0+m_1x+\cdots+m_{k-1}x^{k-1}$, the codeword for $m(x)$ consists of
its evaluations $(m(\alpha_0), \cdots, m(\alpha_{n-1}))$ on 
$n$ distinct field elements $\alpha_0, \cdots, \alpha_{n-1}$, which is received as 
$(\beta_0, \cdots, \beta_{n-1})$.
The GS decoder algorithm is parameterized with a non-negative interpolation multiplicity (order) $\omega\ge 1$.
For each $\omega$, there is an associated decoding radius 
$$t_\omega(n,k)=n-1-\left\lfloor \frac{\max\left\{K:  \displaystyle\sum_{i=0}^{\left\lfloor\frac{K}{k-1}\right\rfloor}({K} -i(k-1) )
\le n{\omega+1\choose 2}\right\}}{\omega}\right\rfloor$$
where we have 
$$t_0(n,k)=\left\lfloor\frac{n-k}{2}\right\rfloor\le t_1(n,k)\le t_2(n,k)\le \cdots \le t_{\omega_0}(n,k)=t_{\omega_0+1}(n,k)=\cdots = t_{GS}(n,k).$$
For a received codeword $(\beta_0, \cdots, \beta_{n-1})$ and an interpolation multiplicity (order) $\omega\ge 1$, 
the GS decoder $GS(\omega)$ finds a list of $L_\omega(n,k)$ polynomials $p_1(x), \cdots, p_{L_\omega(n,k)}(x)$ 
such that one of these polynomials $p_i(x)$ satisfies the condition 
$$\left|\left\{j: p_{i}(\alpha_j)\not=\beta_j\right\}\right|\le t_\omega(n,k)$$
where 
$$L_{\omega}(n,k)=\left\lfloor\sqrt{  \frac{2n{\omega+1\choose 2}}{k-1} + \left(\frac{k+1}{2(k-1)}\right)^2} \right\rfloor -\left(\frac{k+1}{2(k-1)}\right).$$

For a polynomial $Q(x,y)$, we say that $Q(x,y)$ has a zero of multiplicity (order) $\omega$ at $(0,0)$
if $Q(x, y)$ contains no term of total degree less than $\omega$. Similarly, 
we say that $Q(x,y)$ has a zero of multiplicity (order) $\omega$ at $(\alpha,\beta)$
if $Q(x+\alpha, y+\beta)$ contains no term of total degree less than $\omega$.
Note that 
$$\begin{array}{ll}
Q(x+\alpha, y+\beta) &= \displaystyle\sum_{i,j}a_{i,j}(x+\alpha)^i(y+\beta)^j\\
&=\displaystyle\sum_{i,j}a_{i,j}\left(\displaystyle\sum_r{i\choose r}x^r\alpha^{i-r}\right)
\left(\displaystyle\sum_s{j\choose s}y^s\beta^{j-s}\right)\\
&=\displaystyle\sum_{r,s} x^ry^s\displaystyle\sum_{i,j}\left(a_{i,j}{i\choose r} 
{j\choose s}\alpha^{i-r}\beta^{j-s}\right)\\
\end{array}$$
Let $Q_{[r,s]}(\alpha,\beta)= \displaystyle\sum_{i,j}\left(a_{i,j}{i\choose r} 
{j\choose s}\alpha^{i-r}\beta^{j-s}\right)$ be the Hasse derivative. 
Then $Q(x,y)$ has a zero of multiplicity (order) $\omega$ at $(\alpha,\beta)$
if and only if $Q_{[r,s]}(\alpha,\beta)=0$ for all $0\le r+s<\omega$.

The Guruswami-Sudan's (GS) decoding algorithm first constructs a bivariate 
polynomial $Q(x,y)$ such that $Q(x,y)$ has a zero of order $\omega$ at 
each of given pairs $(\alpha_i,\beta_i)$. This could be done by constructing a linear equation
system with $Q(x,y)$'s coeffifients as unknowns. For $Q(x,y)$ to satisfy the requried 
property, it is sufficient to have $Q_{[r,s]}(\alpha_i,\beta_i)=0$ for all $i=0, \cdots, n-1$
and  $r+s<\omega$. That is, we need to solve a linear equation system of 
$O(n\omega^2)$ equations at the cost $O(n^3\omega^6)$ steps.
Specifically, the decoding algorithm $GS(\omega)$  consists of the following two steps.
\begin{enumerate}
\item Constructs a nonzero two-variable polynomial 
$$Q(x,y)=\sum_{i=0}^{n{\omega+1\choose 2}}a_i\phi_i(x,y)$$
where $\phi_0(x,y)<\phi_0(x,y)<\cdots, $ is a list of all monomials $x^iy^j$ ordered by the $(1,k-1)$-lexicographic order.
That is,  $x^{i_1}y^{j_1}<x^{i_2}y^{j_2}$
if and only if ``$i_1+(k-1)j_1< i_2+(k-1)j_2$''  or ``$i_1+(k-1)j_1= i_2+(k-1)j_2$ and $j_1<j_2$''.
The constructed polynomial $Q(x, y)$ satisfies the property that it has a zero of order $\omega$ at each of the
$n$ points $(\alpha_i, \beta_i)$ for $i = 1,\cdots, n$.
\item Factorize the polynomial $Q(x,y)$ to get at most $L_\omega$ univariate polynomials:
$${\cal L}=\left\{p(x): y-p(x)|Q(x,y)\right\}.$$
Among these $L_\omega$ polynomials, one is the transmitted message polynomial $m(x)$.
\end{enumerate}
Note that $Q(x,y)$ has the following properties:
\begin{enumerate}
\item $Q(x,y)$ has at most $n{\omega+1\choose 2}$ terms.
\item The $(1,k-1)$ degree of $Q(x,y)$ is strictly less than $\sqrt{2(k-1)n{\omega+1\choose 2}}$.
\item The $y$-degree of $Q(x,y)$ is at most $L_{\omega}(n,k)$.
\item The $x$-degree of  $Q(x,y)$ is at most $\sqrt{2(k-1)n{\omega+1\choose 2}}$.
\end{enumerate}

Instead of solving a linear equation system for the construction of $Q(x,y)$, 
K\"{o}tter proposed an iterative interpolation 
algorithm to construct the polynomial $Q(x,y)$. In K\"{o}tter's algorithm, one
first defines candidate polynomials $Q_j(x,y)=y^j$ for 
$j=0, \cdots, L_\omega$. Then one recursively revises $Q_j(x,y)$ for each of the 
pairs $(\alpha_i,\beta_i)$ such that $Q_{j,[r,s]}(\alpha_i,\beta_i)=0$ for all
$r+s<\omega$. In case that two of the candidate polynomials $Q_{j_0}(x,y)$
and $Q_{j_1}(x,y)$ do not satisfy this condition for given $r$ and $s$, one revises them as follows:
\begin{itemize}
\item Let $Q_{j_1}(x,y)=Q_{j_0, [r,s]}(\alpha_i,\beta_i) Q_{j_1}(x,y) - Q_{j_1,[r,s]}(\alpha_i,\beta_i) Q_{j_0}(x,y)$.
\item Let $Q_{j_0}(x,y)=Q_{j_0, [r,s]}(\alpha_i,\beta_i)\tilde{Q}_{j_0}(x,y) - \tilde{Q}_{j_0,[r,s]}(\alpha_i,\beta_i) Q_{j_0}(x,y)$
where $\tilde{Q}_{j_0}(x,y)=(x-\alpha_i)Q_{j_0}(x,y)$.
\end{itemize}
Based on the fact that Hasse derivative is bilinear, it follows that, after the above revision, we have 
both $Q_{j_0,[r,s]}(\alpha_i,\beta_i)=0$ and $Q_{j_1,[r,s]}(\alpha_i,\beta_i)=0$.
K\"{o}tter's algorithm runs in time $O(nL_\omega\omega^2Q_{size})=O(n^2\omega^4L_\omega)$ 
where $Q_{size}$ is the number of terms within $Q(x,y)$.

\noindent
{\em Input}: $(\alpha_0, \beta_0), \cdots, (\alpha_{n-1}, \beta_{n-1})$, $\omega$, $L_{\omega}$.

\noindent
{\em Output}: $Q(x,y)$ that has a zero of order $\omega$ at $(\alpha_i,\beta_i)$ for all $i=0, \cdots, n-1$. 

\noindent
{\em Algorithm Steps}:
\begin{enumerate}
\item Let $Q_j(x,y)=y^j$ for $j=0, \cdots, L_{\omega}$.\footnote{For implementation, one may use 
a sparse $\left(1+\sqrt{2(k-1)n{\omega+1\choose 2}}\right)\times\left( 1+L_{\omega}(n,k)\right)$ matrix to denote $Q_j(x,y)$.}
\item For $i=0$ to $n-1$, do the following:
\begin{itemize}
\item For $r=0, \cdots, \omega-1$ do:
\begin{itemize}
\item for $s=0, \cdots, \omega-r-1$ do:
\begin{itemize}
\item Compute Hasse derivative $Q_{j,[r,s]}(\alpha_i,\beta_i)=\sum_{u,v}{u\choose r}{v\choose s}a_{u,v}\alpha_i^{u-r}\beta_i^{v-s}$ at the point
$(\alpha_i, \beta_i)$ for $j=0, \cdots, L_{\omega}$,
where $Q_j(x,y)=\sum_{u,v}a_{u,v}x^uy^v$.
\item Let $J=\{j: Q_{j,[r,s]}(\alpha_i,\beta_i) \not=0\}$. We need to adjust these $Q_j(x,y)$ so that they have a zero
of order $\omega$ at $(\alpha_i, \beta_i)$.
\item If $J\not=\emptyset$, do the following
\begin{itemize}
\item Let $j_0$ be the least index in $J$ such that $Q_{j_0}(x,y)<Q_j(x,y)$ for all $ j\in J$ with the $(1,k-1)$-lexicographic order.
\item For $j\in J$ with $j\not=j_0$, let 
$$Q_j(x,y)=Q_{j_0,[r,s]}(\alpha_i,\beta_i)Q_j(x,y)-Q_{j,[r,s]}(\alpha_i,\beta_i)Q_{j_0}(x,y).$$
\item Let 
$$\begin{array}{ll}
Q_{j_0}(x,y) &= Q_{j_0,[r,s]}(\alpha_i,\beta_i)\tilde{Q}_{j_0}(x,y)-\tilde{Q}_{j_0,[r,s]}(\alpha_i,\beta_i)Q_{j_0}(x,y)\\
&=Q_{j_0, [r,s]}(\alpha_i,\beta_i) xQ_{j_0}(x,y)- \hat{Q}_{j_0,[r,s]}(\alpha_i,\beta_i)  Q_{j_0}(x,y)
\end{array}$$
where $\tilde{Q}_{j_0}(x,y)=(x-\alpha_i)Q_{j_0}(x,y)$  
and $\hat{Q}_{j_0}(x,y)=xQ_{j_0}(x,y)$.
\end{itemize}
\end{itemize}
\end{itemize}
\end{itemize}
\item Let $Q(x,y)=\min\{Q_j(x,y):j\}$ with respect to the $(1,k-1)$-lexicographic order of leading monomials.
\end{enumerate}

The $y$-roots $f(x)=f_0+f_1x+\cdots+f_{k-1}x^{k-1}$ of $Q(x,y)$ could be 
determined by recursively finding the coefficients $f_0, \cdots, f_{k-1}$.
Note that 
\begin{equation}
\label{yrootsequ}
(y-f_0-f_1x-\cdots-f_{k-1}x^{k-1})R(x,y)=Q(x,y)
\end{equation}
for some $R(x,y)$. Thus $(y-f_0)R(0,y)=Q(0,y)$. That is,
$f_0$ is a root of $Q(0,y)$. By substituting $y=xy+f_0$ into
(\ref{yrootsequ}) and then dividing $x^{i_1}$ in both sides
such that $x^{i_1+1}\nmid Q(x,y)$, one obtains
\begin{equation}
\label{yrootsequxy}
\left(y-f_1-f_2x\cdots-f_{k-1}x^{k-2}\right)\frac{R(xy+f_0,y)}{x^{i_1}}=\frac{Q(xy+f_0,y)}{x^{i_1}}
\end{equation}
Thus one has $(y-f_1)R_1(f_0,y)=Q_1(0,y)$ where $R_1(x,y)=\frac{R(xy+f_0,y)}{x^{i_1}}$ and 
$Q_1(x,y)=\frac{Q(xy+f_0,y)}{x^{i_1}}$. That is, $f_1$ is a root of $Q_1(0,y)$.
Continuing this process, one obtains Roth-Ruckenstein factorization algorithm.

\noindent
{\em Input}: $Q(x,y)$, $k-1$. 

\noindent
{\em Output}: all $f(x)$ of degree at most $k-1$ such that $(y-f(x))| Q(x,y)$.

\noindent
{\em Algorithm Steps}:
\begin{enumerate}
\item Let $\pi[0]=\mbox{NULL}$, $\deg(0)=-1$, $Q_0(x,y)=Q(x,y)$, $t=1$, and $u=0$.
\item Run the depth-first search DFS$(u)$ where DFS$(u)$ is defined as:
\begin{itemize}
\item If $Q_u(x,0)=0$, output $f^u(x)=f^u_{\deg(u)}x^{\deg(u)}+f^{u_0}_{\deg(u_0)}x^{\deg(u_0)}+f^{u_1}_{\deg(u_1)}x^{\deg(u_1)}+\cdots$ where $u_0$ is the parent of $u$, $u_1$ is the parent of $u_0$, and so on.
\item If $Q_u(x,0)\not=0$ and $\deg(u)<k-1$ then do the following:
\begin{itemize}
\item For each root $\alpha$ of $Q_u(0,y)$ do:
\begin{itemize}
\item Let $v=t, t=t+1$;
\item $\pi[v]=u, \deg(v)=\deg(u)+1$, $f^v_{\deg{v}}=\alpha$, 
\item $Q_v(x,y)=\frac{Q_u(x,y)}{x^i}$ such that $x^i| Q_u(x,y)$ but $x^{i+1}\nmid Q_u(x,y)$.
\item Do DFS$[v]$.
\end{itemize}
\end{itemize}
\end{itemize}
\end{enumerate}
In the above algorithm, we have the following notations:
\begin{itemize}
\item $\pi[u]$ is the parent of $u$
\item $\deg(u)$ is the degree of $u$. That is, the distance from root minus 1.
\item $f^u_{\deg(u)}$ is the polynomial coefficient at $x^{\deg(u)}$.
\end{itemize}

In the above Roth-Ruckenstein algorithm, we need to compute all roots of $Q_u(0,y)$. This could be done using 
any of the root-finding algorithms discussed in preceding sections. For example, one may use exhaustive search, Chien's search,
Berlekamp Trace Algorithm (BTA), or equal-degree factorization by Cantor and Zassenhaus.
In the above Roth-Ruckenstein algorithm, we also need to compute $Q(x,xy+\alpha)$ from $Q(x,y)=\sum_{i,j}a_{i,j}x^iy^j$.
Note that 
$$\begin{array}{ll}
Q(x, xy+\alpha) &=\displaystyle\sum_{r,j}a_{r,j}x^r(xy+\alpha)^j\\
&=\displaystyle\sum_{r,j}a_{r,j}x^r\left(\displaystyle\sum_s{j\choose s}x^sy^s\alpha^{j-s}\right)\\
&=\displaystyle\sum_s\left(\displaystyle\sum_{r,j}a_{r,j}{j\choose s}\alpha^{j-s}x^{r+s}y^s\right)\\
&=\displaystyle\sum_{r,s}\left(x^{r+s}y^s\displaystyle\sum_{j}a_{r,j}{j\choose s}\alpha^{j-s}\right)\\
&=\displaystyle\sum_{r,s} Q_{r,s}(\alpha)x^{r+s}y^s
\end{array}$$
where 
$$Q_{r,s}(y)=\sum_{j\ge s}{j\choose s}a_{r,j}y^{j-s}.$$

Several more efficient interpolation/factorization algorithms for list decoding have been proposed in the last decades, for example,
\cite{alekhnovich2002linear,beelen2013rational,chowdhury2015faster,nielsen2015power,trifonov2010efficient,zeh2011interpolation}.
Our experiments show that they are still quite slow for RLCE encryption scheme. Thus the advantages of reducing key sizes by using 
list-decoding may be limited for RLCE schemes.

\subsection{Experimental results}
\begin{table}[th]
\caption{Milliseconds for decoding Reed-Solomon codes over $GF(2^{m})$}
\label{rsper}
\begin{center}
$\begin{array}{|c|c|c|}\hline
(n,k,t,m)& \mbox{BM-decoder}& \mbox{Euclidean decoder}  \\ \hline
(532,376,78,10)&1.8763225&2.6413376\\ \hline
(630,470,80,10)&1.9261904&2.6511796\\ \hline
(846,618,114,10)&3.0183825&3.6363407 \\ \hline
(1000,764,118,10)&3.1226213&4.0247824\\ \hline
(1160,700,230,11)&10.3142787&13.3073421\\ \hline
(1360,800,280,11)&12.4488992&16.3140049\\ \hline
\end{array}$
\end{center}
\end{table}
Table \ref{rsper} gives experimental results
on decoding Reed-Solomon codes for various parameters corresponding RLCE schemes. The implementation was run on a MacBook Pro
with masOS Sierra version 10.12.5 with 2.9GHz Intel Core i7 Processor. The reported time is the required milliseconds 
for decoding a received codeword over $GF(2^m)$ (an average of 10,000 trials).

For the list-decoding based RLCE encryption scheme, we tested Reed-Solomon codes with
$(n,k,t,\omega, L_\omega, m)=(520,380,73,9,10,10))$. It takes 1865 seconds (that is, approximately 31 minutes)
to decode a received code.

\section{Conclusion}
This paper compares different algorithms for implementing the RLCE encryption scheme. The experiments show that for all of the 
RLCE encryption scheme parameters (corresponding to AES-128, AES-192, and AES-256), Chien's search algorithm should be used 
in the root-finding process of the error locator polynomials. For list-decoding based RLCE schemes, the root-finding process for small degree 
polynomials should use BTA algorithm for polynomial degrees smaller than 5 and Chien's search for polynomial degrees above 5.
For polynomial multiplications, one should use optimized classical polynomial multiplicaton algorithm for polynomials of degree 115 and less.
For polynoials of degree 115 and above, one should use Karatsuba algorithm. For matrix multiplications, one should use optimized classical
matrix multiplicaiton algorithm for matrices of dimension 750 or less. For matrices of dimension 750 or above, one should use Strassen's algorithm.
For the underlying Reed-Solomon decoding process, Berlekamp-Massey outperforms Euclidean decoding process. 

\bibliographystyle{plain}


\begin{thebibliography}{10}

\bibitem{alekhnovich2002linear}
M.~Alekhnovich.
\newblock Linear diophantine equations over polynomials and soft decoding of
  reed-solomon codes.
\newblock In {\em Proc. 43rd IEEE FOCS}, pages 439--448. IEEE, 2002.

\bibitem{bard2006accelerating}
G.V. Bard.
\newblock Accelerating cryptanalysis with the method of four russians.
\newblock {\em IACR Cryptology EPrint Archive}, 2006:251, 2006.

\bibitem{beelen2013rational}
P.~Beelen, T.~H{\o}holdt, J.S.R. Nielsen, and Y.~Wu.
\newblock On rational interpolation-based list-decoding and list-decoding
  binary goppa codes.
\newblock {\em IEEE Tran. Information Theory}, 59(6):3269--3281, 2013.

\bibitem{berlekamp1968algebraic}
E.R. Berlekamp.
\newblock {\em Algebraic coding theory}.
\newblock McGraw-Hill, 1968.

\bibitem{bernstein2008attacking}
D.J. Bernstein, T.~Lange, and C.~Peters.
\newblock Attacking and defending the {McEliece} cryptosystem.
\newblock In {\em Proc. Int. Workshop PQC}, pages 31--46. Springer, 2008.

\bibitem{bunch1974triangular}
J.R. Bunch and J.E. Hopcroft.
\newblock Triangular factorization and inversion by fast matrix multiplication.
\newblock {\em Mathematics of Computation}, 28(125):231--236, 1974.

\bibitem{cantor1989arithmetical}
D.G. Cantor.
\newblock On arithmetical algorithms over finite fields.
\newblock {\em Journal of Combinatorial Theory, Series A}, 50(2):285--300,
  1989.

\bibitem{chowdhury2015faster}
M.F.I. Chowdhury, C.-P. Jeannerod, V.~Neiger, E.~Schost, and G.~Villard.
\newblock Faster algorithms for multivariate interpolation with multiplicities
  and simultaneous polynomial approximations.
\newblock {\em IEEE Tran. Information Theory}, 61(5):2370--2387, 2015.

\bibitem{gao2010additive}
S.~Gao and T.~Mateer.
\newblock Additive fast fourier transforms over finite fields.
\newblock {\em IEEE Tran. Information Theory}, 56(12):6265--6272, 2010.

\bibitem{guruswami1998improved}
V.~Guruswami and M.~Sudan.
\newblock Improved decoding of {Reed-Solomon} and algebraic-geometric codes.
\newblock {\em IEEE Tran. Information Theory}, 45:1757--1767, 1999.

\bibitem{kotter1996fast}
R.~K\"{o}tter.
\newblock Fast generalized minimum-distance decoding of algebraic-geometry and
  {Reed-Solomon} codes.
\newblock {\em IEEE Tran. Information Theory}, 42(3):721--737, 1996.

\bibitem{massey1969shift}
J.~Massey.
\newblock Shift-register synthesis and bch decoding.
\newblock {\em IEEE Trans. Information Theory}, 15(1):122--127, 1969.

\bibitem{moenck1976practical}
R.T. Moenck.
\newblock Practical fast polynomial multiplication.
\newblock In {\em Proc. 3rd ACM Symposium on Symbolic and algebraic
  computation}, pages 136--148. ACM, 1976.

\bibitem{nielsen2015power}
J.S.R. Nielsen.
\newblock Power decoding {Reed--Solomon} codes up to the {Johnson} radius.
\newblock {\em arXiv preprint arXiv:1505.02111}, 2015.

\bibitem{roth2000efficient}
R.M. Roth and G.~Ruckenstein.
\newblock Efficient decoding of reed-solomon codes beyond half the minimum
  distance.
\newblock {\em IEEE Trans. Information Theory}, 46(1):246--257, 2000.

\bibitem{sudan1997decoding}
M.~Sudan.
\newblock Decoding of {Reed-Solomon} codes beyond the error-correction bound.
\newblock {\em J. complexity}, 13(1):180--193, 1997.

\bibitem{trifonov2010efficient}
P.V. Trifonov.
\newblock Efficient interpolation in the guruswami--sudan algorithm.
\newblock {\em IEEE Tran. Information Theory}, 56(9):4341--4349, 2010.

\bibitem{von1996arithmetic}
J.~Von~zur Gathen and J.~Gerhard.
\newblock Arithmetic and factorization of polynomial over f 2.
\newblock In {\em Proc. ISSAC}, pages 1--9. ACM, 1996.

\bibitem{7541753}
Y.~Wang.
\newblock Quantum resistant random linear code based public key encryption
  scheme {RLCE}.
\newblock In {\em Proc. IEEE ISIT}, pages 2519--2523, July 2016.

\bibitem{wangrlcelong}
Y.~Wang.
\newblock Revised quantum resistant public key encryption scheme {RLCE} and
  {IND-CCA2} security for {McEliece} schemes.
\newblock In {\em IACR ePrint https://eprint.iacr.org/2017/206.pdf}, July 2017.

\bibitem{zeh2011interpolation}
A.~Zeh, C.~Gentner, and D.~Augot.
\newblock An interpolation procedure for list decoding reed--solomon codes
  based on generalized key equations.
\newblock {\em IEEE Tran. Information Theory}, 57(9):5946--5959, 2011.

\end{thebibliography}

\end{document}